\documentclass[twocolumn,aps,prd,amsmath,amssymb,nofootinbib]{revtex4-1}\newcommand{\cw}{\columnwidth}\newcommand{\htt}{7.2cm}\newcommand{\htb}{7.8cm}\newcommand{\htz}{7.5cm}\newcommand{\preprintnumber}{\hfill MIT-CTP 4571\,\,\,\,\maketitle}

\usepackage{graphicx,bm,enumerate}

\newcommand{\beq}{\begin{equation}}
\newcommand{\bea}{\begin{eqnarray}}
\newcommand{\eeq}{\end{equation}}
\newcommand{\eea}{\end{eqnarray}}
\newcommand{\mpl}{M_{Pl}}
\newcommand{\Vp}{\tilde{V}}
\newcommand{\vp}{\tilde{v}}
\newcommand{\Tp}{\tilde{t}}
\newcommand{\phip}{\tilde{\phi}}
\newcommand{\ed}{\varepsilon}
\newcommand{\edp}{\tilde{\varepsilon}}
\newcommand{\pr}{p}
\newcommand{\pp}{\tilde{p}}
\newcommand{\nd}{n}
\newcommand{\phis}{\xi}
\newcommand{\ex}{q}
\newcommand{\F}{\Lambda}
\newcommand{\Pv}{r}

\newcommand{\phim}{\rho}

\newcommand{\psib}{\zeta}
\newcommand{\N}{\mathcal{N}}
\newcommand{\Num}{N}

\newcommand{\bet}{\beta}

\newcommand{\kin}{X}

\newcommand{\lamh}{\hat\lambda}

\begin{document}

\title{A THEORY OF SELF-RESONANCE AFTER INFLATION\\
\vspace{0.1cm}
Part 1: Adiabatic and Isocurvature Goldstone Modes}

\begin{abstract}
We develop a theory of self-resonance after inflation. We study a large class of models involving multiple scalar fields
with an internal symmetry. For illustration, we often specialize to dimension 4 potentials, but we derive results for general potentials. This is the first part of a two part series of papers. Here in Part 1 we especially focus on the behavior of long wavelengths modes, which are found to govern most of the important physics. Since the inflaton background spontaneously breaks the time translation symmetry and the internal symmetry, we obtain Goldstone modes; these are the adiabatic and isocurvature modes. We find general conditions on the potential for when a large instability band exists for these modes at long wavelengths. For the adiabatic mode, this is determined by a sound speed derived from the time averaged potential. While for the isocurvature mode, this is determined by a speed derived from a time averaged auxiliary potential. Interestingly, we find that this instability band usually exists for one of these classes of modes, rather than both simultaneously. We focus on backgrounds that evolve radially in field space, as setup by inflation, and also mention circular orbits, as relevant to Q-balls. 
In Part 2 \cite{Part2} we derive the central behavior from the underlying description of many particle quantum mechanics, 
and introduce a weak breaking of the symmetry to study corrections to particle-antiparticle production from preheating.
\end{abstract}

\author{
Mark P.~Hertzberg$^*$, 
Johanna Karouby$^\dagger$, 
William G.~Spitzer,
Juana C.~Becerra, 
Lanqing Li
}
\affiliation{Center for Theoretical Physics and Dept.~of Physics,\\ 
Massachusetts Institute of Technology, Cambridge, MA 02139, USA}

\let\thefootnote\relax\footnotetext{$^*$Electronic address: {\tt mphertz@mit.edu}\\$^\dagger$Electronic address: {\tt karoubyj@mit.edu}}

\date{\today}

\preprintnumber

\maketitle

\tableofcontents

\section{Introduction} \label{Introduction}

Cosmological inflation is a successful theory of the early universe \cite{Guth:1980zm,Linde:1981mu,Albrecht:1982wi}, which accounts for the approximately scale-invariant distribution of structures on large scales. The structures arises from quantum fluctuations in the inflaton scalar field/s $\phi$ by stretching modes to large scales due to inflation's exponential expansion. 
Evidence for inflation is increasing with detailed measurement of the cosmic microwave background radiation \cite{WMAP,Planck,BICEP,Dodelson:2009kq}, and this has motivated the construction of many interesting theoretical models \cite{ Lyth:1998xn,Kachru:2003sx,Dimopoulos:2005ac,Linde:2007fr,Hertzberg:2011rc,Hertzberg:2014aha,Baumann:2014nda,Hertzberg:2014sza,Kaiser:2013sna}.

Once inflation has ended, the quantum fluctuations are no longer exponentially stretched. However, another interesting phenomenon can sometimes come into play. As the inflaton oscillates back and forth in its potential, it can cause quantum fluctuations to grow rapidly; an example of parametric resonance.
For inflationary models, there can be resonance in daughter fields (preheating) or to self-resonance in the inflaton field itself. Often in the literature the focus has been on coupling to daughter fields. 

Various interesting and important work includes Refs.~\cite{Kofman:1994rk,Kofman:1997yn,Greene:1997fu,Battefeld:2008bu,Greene:1997ge,Bassett:1999ta,Dufaux:2006ee,Greene:1998nh,Greene:2000ew,Peloso:2000hy,Davis:2000zp,Braden:2010wd,Deskins:2013dwa,Barnaby:2011qe,GarciaBellido:2008ab,Figueroa:2009jw,Ashoorioon:2013oha,Amin:2010dc,Gleiser:2011xj,Amin:2011hj,Hertzberg:2010yz}. 
For example, classic work \cite{Kofman:1994rk,Kofman:1997yn} emphasized a coupling of the inflaton $\phi$ to a daughter field $\chi$, with interactions such as $\sim g^2\phi^2\chi^2$ or $\sim g\,\phi\,\chi^2$, which can cause explosive growth in $\chi$ for some parameter regimes. Other important possibilities include coupling to gauge fields \cite{Davis:2000zp,Deskins:2013dwa}, abelian or non-abelian, coupling to fermonic fields \cite{Greene:1998nh,Greene:2000ew}, metric preheating \cite{Bassett:1999ta}, and so on.
On the other hand, self-resonance can occur for potentials with nonlinearities, including the quartic term $\sim\lambda\,\phi^4$, as discussed in \cite{Greene:1997fu}. In fact this can lead to coherent structures, such as oscillons, for negative $\lambda$; see \cite{Amin:2010dc,Gleiser:2011xj,Amin:2011hj}.
Here we focus on the important issue of self-resonance of the inflaton, and assume couplings to other fields are small. 
We will re-organize the analysis of self-resonance into a kind of fluid description for long wavelengths, which appears to go beyond the existing literature.

An important question is whether this self-resonance is efficient, i.e., whether it causes significant resonance for some range of $k$-modes. If so, this can provide a corresponding enhancement in the power spectrum and possible fragmentation of the inflaton field. Another important question is whether these modes are only on very small sub-Hubble scales, as is usually thought to be the case in the post-inflationary era, or whether there can be some enhancement for order Hubble or super-Hubble scales. In some of the simplest models of single field inflation, such as $\sim \lambda\,\phi^4$, the answer to both of these questions is in the negative, i.e., the resonance is rather inefficient \cite{Greene:1997fu} and is usually restricted to highly sub-Hubble modes.

This is Part 1 of a series of two papers. In these papers we consider multi-field inflation models and more general potentials. This much more general framework is motivated from the point of view of high energy physics, as occurs in frameworks such as supersymmetry, string theory, and beyond. For simplicity, we consider the potential $V$ to carry an internal $\mathcal{O}(\N)$ symmetry, so it is only a function of $|\vec\phi|$; this may be required by a gauge redundancy or, more likely, due to an approximate global symmetry. We show that in this class of models, the resonance is often efficient and is predominantly given by somewhat long wavelengths. We then investigate the conditions under which a large instability band exists for long wavelengths in the post-inflationary era.

In the presence of multiple fields, there are correspondingly multiple modes of excitations that can potentially be resonant. In this first paper, we decompose these modes into the {\em adiabatic} and {\em isocurvature} modes of the inflaton and discuss under what conditions either of them has significant resonance for long wavelengths. We show that these modes exhibit a gapless spectrum, as required by the Goldstone theorem. We show that the existence of an instability in the adiabatic mode can be derived from a sound speed associated with the pressure and density of the background, while the existence of an instability in the isocurvature mode can be derived from a speed associated with the pressure and density of an auxiliary background that we define. We find that the isocurvature mode can lead to enhanced power on especially long wavelengths.

Finally, we study the case of two fields, and organize the inflaton into a complex scalar. 
For background circular motion in the complex plane, we derive the conditions for breakup towards so called Q-balls. 

In Part 2 \cite{Part2}, we show that for potentials that give rise to an unstable isocurvature mode, the inflaton fragments into regions of particles and antiparticles. As an application, we connect our analysis to inflationary baryogenesis models as formulated by some of us recently in Refs.~\cite{Hertzberg:2013jba,Hertzberg:2013mba}. In particular, we allow for a small breaking of the internal symmetry and derive corrections to the particle asymmetry, which may be responsible for the late time baryon asymmetry.

The outline of this paper is as follows: 
In Section \ref{Model} we present the class of models under investigation and outline its equations of motion and Floquet theory. 
In Section \ref{Quartic} we numerically solve the problem for dimension 4 potentials built out of a quadratic mass term and a quartic interaction term.
In Section \ref{AuxiliaryPotential} we derive a type of auxiliary potential that controls the behavior of the isocurvature modes. 
In Section \ref{LongWavelength} we derive the general conditions on the potential for when an instability exists at long wavelengths.
In Section \ref{Circular} we explore backgrounds that are circular in the complex field plane.
In Section \ref{Conclusions} we discuss our findings and conclude. 
Finally, in Appendix \ref{NonCanonical} we generalize the analysis to non-canonical kinetic terms.

\section{Symmetric Theories} \label{Model} 

Inflation is a theory of the early universe driven by one or more scalar fields coupled to gravity.
Let us consider $\N$ scalar fields. For convenience, we organize them into a vector
\beq
\vec{\phi}=\{\phi_1,\ldots,\phi_\N\}
\eeq
In the case of two scalar fields, it is often useful to organize $\phi$ into a complex scalar as follows
\beq
\phi = {\phi_1+i\phi_2\over\sqrt{2}}
\eeq
We shall focus on this complex field later, but more generally we shall focus on an arbitrary number of fields $\N$.

The inflationary action is, in general, some effective field theory, since gravitation is known to be non-renormalizable in 4 dimensions.
A reasonable assumption is that all higher order derivative corrections are suppressed by a sufficiently large mass scale that they can be ignored. This allows us to simply focus on the leading order two-derivative action. The most general form of the action may then be written as (signature $-+++$, units $\hbar=c=1$)
\beq
S=\int d^4x\sqrt{-g}\left[{\mpl^2\over2}\mathcal{R}-{1\over2}G_{ij}(\vec\phi)\partial_\mu\phi^i\partial^\mu\phi^j-V(\vec\phi)\right]\,\,
\eeq
where $\mpl\equiv1\sqrt{8\pi G_N}$ is the reduced Planck mass. Here we have expressed the action, without loss of generality, in the Einstein frame where the gravity sector is canonical. The matrix $G_{ij}$ is the metric on field space, which in general leads to a type of non-linear sigma model. In Appendix \ref{NonCanonical} we consider general forms for the kinetic energy. For now we restrict attention to canonical kinetic energy with
\beq
G_{ij}=\delta_{ij}
\eeq
We note that this approximation is technically natural. That is, if we assume the kinetic term is canonical, we find that the quantum corrections tend to be small. The reason is that this form of the kinetic term is broken only by interactions between the fields in the potential sector, which are suppressed by the strength of the couplings; these couplings are typically small to achieve models of inflation with $\sim 10^{-10}$ level variance in fluctuations, though there can be exceptions. 

Our freedom then lies in the choice of the potential $V(\vec\phi)$. For simplicity we consider models that carry an internal rotational symmetry
\beq
\phi^i\to R^i_j\,\phi^j
\eeq 
where $R$ is a rotation matrix acting on field space. Formally this implies an $O(\N)$ symmetry and the potential may be written as $V(\vec\phi)=V(|\vec\phi|)$. This group of symmetries may, for instance, be enforced by a gauge redundancy (``gauge symmetry"). It is non-trivial, however, to charge the inflaton since one then needs to explain why the inflaton's self interactions are small enough to ensure the $\sim 10^{-10}$ level variance in fluctuations; though it is conceivable. Another possibility is simply to appeal to an approximate global symmetry. 
In Part 2 \cite{Part2} we introduce a small breaking of this global symmetry and show how to utilize results from the symmetric theory to obtain the leading order corrections to the generation of particle number and baryogenesis.

\subsection{Background Evolution}\label{Background}

Inflation inevitably forces the background space-time to a flat FRW metric
\beq
ds^2 = -dt^2 +a(t)^2 d{\bf x}^2
\eeq
where $a(t)$ is the scale factor. Furthermore, in the slow-roll phase of inflation, any angular motion in field space will be redshifted away. This results in essentially radial motion in field space. This is an attractor solution of inflation due to the internal symmetry in field space. 

This purely radial motion for the background shall be denoted by the field $\phi_0(t)$. We can, without loss of generality, orient our field space co-ordinates, such that the background points along the $\phi_\N$ direction, i.e., 
\beq
\vec\phi_0(t) = \{0,\ldots,0,\phi_0(t)\}
\label{BgdDirection}\eeq
This background field satisfies the equation of motion
\beq
\ddot\phi_0+3H\dot\phi_0+V'(\phi_0) = 0
\eeq
where $H=\dot a/a$ is the Hubble parameter. During slow-roll inflation, the second and third terms here dominate. After inflation, as is the focus of this work, the first and third terms dominate and the second ``friction" term is sub-dominant.

\subsection{Linearized Perturbations} \label{Linearized}

We shall denote the $\phi_\N$ direction as ``parallel" since it is parallel to the background. The other $\N-1$ directions shall be denoted as ``orthogonal" since they are orthogonal to the background. We can then expand the field around the background as 
\beq
\vec\phi({\bf x},t) = \vec\phi_0(t)+\delta\vec\phi({\bf x},t)
\eeq
where
\bea
\delta\vec\phi({\bf x},t) = \{\delta\phi_{\perp 1}({\bf x},t),\ldots,\delta\phi_{\perp \N-1}({\bf x},t),\delta\phi_\parallel({\bf x},t)\}\,\,\,\,\,\,
\eea
Expanding the scalar field equations to linear order, the equations of motion for these perturbations are found to be
\bea
&&\,\ddot{\delta\phi_\parallel}+3H\dot{\delta\phi_\parallel}+\left({k^2\over a^2}+V''(\phi_0)\right)\delta\phi_\parallel=\mathcal{G}\label{PertEqn1}\\
&&\ddot{\,\delta\phi_{\perp i}}+3H\!\dot{\,\delta\phi_{\perp i}}+\left({k^2\over a^2}+{V'(\phi_0)\over\phi_0}\right)\delta\phi_{\perp i}=0\label{PertEqn2}
\eea
where we have Fourier transformed to $k$-space. For the orthogonal components, we have included an ``$i$" index, where $i$ runs over $i=1,\ldots,\N-1$; each equation carries the same structure due to the symmetry. If we ignore linear corrections to the metric, then we have $\mathcal{G}=0$, and the right hand side of eq.~(\ref{PertEqn1}) becomes trivial. Otherwise, we can include linear corrections to the metric, whose form depends on gauge. For example, one can work in a gauge with flat hypersurfaces, and one finds that local gravity gives rise to the following correction on the right hand side \cite{Nakamura:1996da}
\beq
\mathcal{G}={1\over a^3\mpl^2}{d\over dt}\!\left(a^3\dot\phi_0^2\over H\right)\delta\phi_\parallel
\label{PertEqn1cor}
\eeq
We will comment on corrections from local gravity further in Part 2 \cite{Part2}. We note that there are no such corrections from local gravity to the orthogonal modes in eq.~(\ref{PertEqn2}); this is associated with the fact that these are {\em isocurvature} modes, as we will explain later.

\subsection{Floquet Theory for Self-Resonance} \label{Floquet}

The above set of equations can, in principle, be directly solved numerically. However a tremendous amount of analytical and semi-analytical progress can be made with an appropriate simplification, as we now describe.

After inflation has ended, the Hubble friction term becomes sub-dominant to the other terms in these equations. This is true for both the background equation and the perturbation equations. The Hubble term is then primarily responsible for a type of slow redshifting of the fields. This effect shall be incorporated in Part 2 \cite{Part2}. For now, we shall focus on time scales short compared to the Hubble time. On these time scales, the background field $\phi_0$ oscillates rapidly back and forth in the potential $V$. This means that to a good approximation $\phi_0(t)$ is {\em periodic}. Furthermore, in the eqs.~(\ref{PertEqn1},\,\ref{PertEqn2}) the background will provide a periodic pump for the perturbations through the terms $V''(\phi_0(t))$ and $V'(\phi_0(t))/\phi_0(t)$. 

In fact an approximate way to handle the expansion is to re-scale the fields by defining $\tilde\phi\equiv a^{3/2}\phi$, which captures the overall red-shifting of the field. The effective mass can be re-defined appropriately, although the final result still carries some mile red-shift dependence. With this in mind, the essential physics is captured by formally sending $H\to 0$ and $a\to 1$, and then each of the perturbation equations become a form of {\em Hill's equation}. We will reinstate these red-shifting effects in Part 2 \cite{Part2}.
So we are led, to good approximation, to a form for the perturbations as
\beq
\ddot{\delta\phi}+h(t)\delta\phi = 0
\eeq
where $h(t)$ is the appropriate periodic pump. In our cases of interest it is given by
\beq
h(t) = \Bigg{\{}
\begin{array}{c}
 k^2 + V''(\phi_0(t))\,\,\,\,\mbox{for}\,\,\,\delta\phi_\parallel\\
 k^2 + {V'(\phi_0(t))\over\phi_0(t)}\,\,\,\,\,\,\,\,\mbox{for}\,\,\,\delta\phi_\perp
\end{array}
\eeq

It is convenient to write the second order equation of motion as a pair of first order equations of motion. To do this, let's define $\delta\pi\equiv\dot{\delta\phi}$. Then Hill's equation becomes
\beq
{d\over dt}\left(\begin{array}{c}
\delta\phi\\
\delta\pi
\end{array}\right)
=\left(\begin{array}{cc}
0 & 1\\
-h(t) & 0
\end{array}\right)
\left(\begin{array}{c}
\delta\phi\\
\delta\pi
\end{array}\right)
\eeq
According to Floquet theory, the late time behavior of this system is determined by the eigenvalues of a certain matrix that we now describe. 

Firstly, a complete basis of solutions comes from considering the following sets of initial conditions
\beq
\left(\begin{array}{c}
\delta\phi(t_i)\\
\delta\pi(t_i)
\end{array}\right)
=\left(\begin{array}{c}
1\\
0
\end{array}\right),\,\,\,\,\,\,
\left(\begin{array}{c}
\delta\phi(t_i)\\
\delta\pi(t_i)
\end{array}\right)
=\left(\begin{array}{c}
0\\
1
\end{array}\right)
\eeq
We organize this space of initial conditions into a matrix. We then numerically evolve this matrix through one period $T$ of the background pump, giving an output matrix, we call $M$. The evolution through $n$ periods is then determined by the matrix $M^n$. For arbitrary initial conditions, the solution after time $t$ (assuming $t$ is an integer multiple of the period $T$) is given by
\beq
\left(\begin{array}{c}
\delta\phi(t)\\
\delta\pi(t)
\end{array}\right)
=M^{t/T}\left(\begin{array}{c}
\delta\phi(t_i)\\
\delta\pi(t_i)
\end{array}\right)
\eeq
The matrix $M$ can be diagonalized in the standard way
\beq
M = S \,D\, S^{-1}
\eeq
where $S$ is a matrix formed from the eigenvectors of $M$, and $D$ is a diagonal matrix comprised of the eigenvalues $\lambda_1,\lambda_2$ of $M$. One can prove that the determinant of $M$ must be 1, so $\lambda=\lambda_1=1/\lambda_2$. The evolution can then be written as
\beq
\left(\begin{array}{c}
\delta\phi(t)\\
\delta\pi(t)
\end{array}\right)
=S\left(\begin{array}{cc}
e^{\mu_k t} & 0\\
0 & e^{-\mu_k t}
\end{array}\right)S^{-1}
\left(\begin{array}{c}
\delta\phi(t_i)\\
\delta\pi(t_i)
\end{array}\right)
\eeq
where we have written the time dependence in terms of an exponential $\sim \exp(\mu_k t)$, where $\mu_k$ is the so called Floquet exponent
\beq
\mu_k = {1\over T}\log(\lambda)
\eeq
We have introduced a $k$ subscript to indicate that the value of the Floquet exponent depends on wavenumber.
If the real part of $\mu_k$ is non-zero, then there is exponential growth of perturbations. Otherwise, if $\mu_k$ is purely imaginary, then there is only oscillatory, or stable, evolution of perturbations.

\section{Motivation from Dimension 4 Potentials} \label{Quartic}

Let us begin by considering the regime well after inflation where the potential is well approximated by its leading order operators. Since the potential is assumed to carry an internal rotational symmetry, we can expand it as
\beq
V(\vec\phi) = V_0 + {1\over 2}m^2|\vec\phi|^2 +{\lambda\over 4}|\vec\phi|^4+\ldots
\eeq
For sufficiently small field amplitudes, these leading dimension 4 terms will dominate the dynamics. Such a regime will normally arise after a sufficient amount of redshifting has occurred. A counter example would be if some of the above coefficients happen to vanish; we will consider this possibility in Part 2 \cite{Part2}. For large amplitudes, higher order corrections to the potential may be important (we mention some examples in Section \ref{PurePowerAdi}).

We will explore the various possibilities, including $\lambda>0$ and $\lambda<0$. In the latter case, higher order terms are necessarily important to ensure stability of the potential relevant for inflation. We will normally focus on a regular mass term $m^2>0$, but will discuss the tachyonic case $m^2<0$ also. The constant term $V_0$ will be chosen to ensure the vacuum energy is zero (the late time dark energy is irrelevant in this early era). So for $m^2>0$, we choose $V_0=0$, and for $m^2<0$, we choose $V_0>0$. 

For now we truncate the potential to purely dimension $\leq 4$ terms and numerically solve for the corresponding Floquet exponent using the method of Section \ref{Floquet}.

\begin{figure*}[t]
  \includegraphics[width=\textwidth]{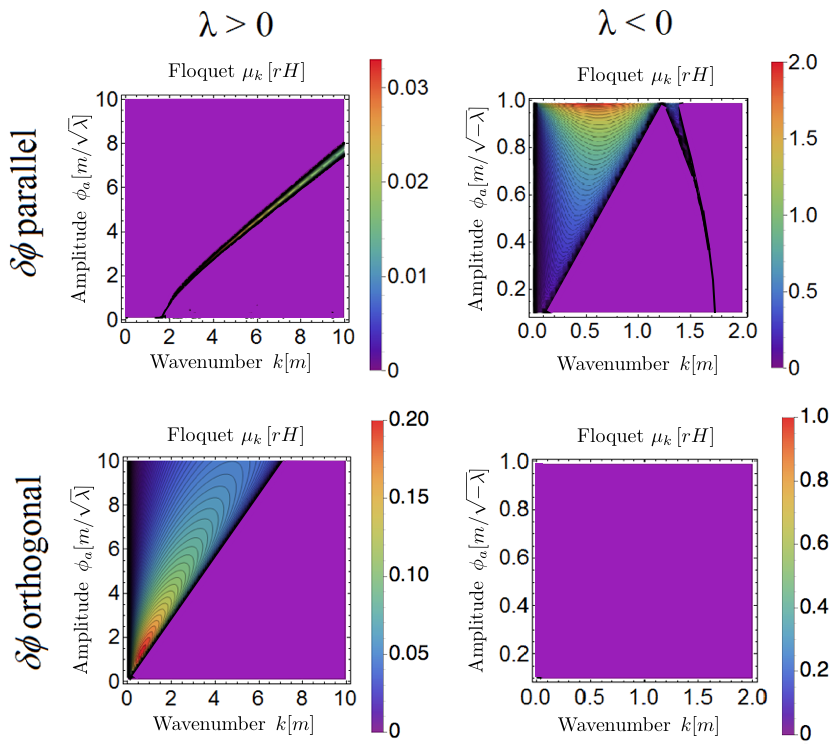}
  \caption{Contour plot of the real part of Floquet exponent $\mu_k$ for dimension 4 potentials as a function of wavenumber $k$ and background amplitude $\phi_a$ with $m^2>0$. Left panel is $\lambda>0$ and right panel is $\lambda<0$. Upper panel is $\delta\phi_\parallel$ and lower panel is $\delta\phi_{\perp}$. We have plotted $\mu_k$ in units of $\Pv H$ where $\Pv\equiv \sqrt{|\lambda|}\,\mpl/m$, $k$ in units of $m$, and $\phi_a$ in units of  $m/\sqrt{|\lambda|}$.}
  \label{4PanelView}\end{figure*}

\subsection{Positive vs Negative Quartic Behavior}

We begin by considering a regular mass term $m^2>0$, so the potential's minimum is at $\vec\phi=0$. We compare the cases in which the quartic coupling $\lambda$ is either positive or negative.

The background field $\phi_0(t)$ evolves under the equation of motion
\beq
\ddot\phi_0+m^2\phi_0+\lambda\,\phi_0^3 = 0
\eeq
This oscillates with some amplitude $\phi_a$. We note that in the case in which $\lambda<0$, the potential exhibits a hill-top, so there is a maximum amplitude. This is given by $\phi_{a,max}=m/\!\sqrt{|\lambda|}$. A natural dimensionless measure of the amplitude is
\beq
\Phi_a\equiv{\phi_a \sqrt{|\lambda|}\over m}
\eeq
with $\Phi_{a,max}=1$ when $\lambda<0$.

The linearized perturbations solve Hill's equation with $h$ function
\beq
h(t) = \Bigg{\{}
\begin{array}{c}
 k^2 + m^2+3\,\lambda\,\phi_0^2(t)\,\,\,\,\mbox{for}\,\,\,\delta\phi_\parallel\\
 k^2 + m^2+\lambda\,\phi_0^2(t)\,\,\,\,\,\,\,\,\mbox{for}\,\,\,\delta\phi_\perp
\end{array}
\label{HillFunction}\eeq
We have numerically solved for the corresponding Floquet exponents, with results for the real part of $\mu_k$ given in Fig.~\ref{4PanelView}.

In the left hand panel we have $\lambda>0$ and in the right hand panel we have $\lambda<0$. In these contour plots we have rescaled the Floquet exponent $\mu_k$ to a certain dimensionless quantity involving Hubble and the Plank mass $\mpl$. 
Although we have ignored Hubble in the analysis, it is still useful to rescale $\mu_k$ by its value. We will discuss these details in Part 2 \cite{Part2}.
The Hubble parameter $H$ is given from the Friedmann equation as
\beq
H^2 = {1\over 3\mpl^2}V(\phi_a)
\eeq
where we have evaluated the energy density at the amplitude of an oscillation, which is therefore purely given by the potential energy.
This naturally introduces the Planck scale, which for the present purposes we would like to scale out. As we discuss in Part 2 \cite{Part2}, the dimensionless parameter that controls the amount of resonance in the problem is
$\Pv\equiv \sqrt{|\lambda|}\mpl/m$.
In Fig.~\ref{4PanelView} we plot the variable $\mu_k/(\Pv H)$, which scales out all physical parameters in the problem.

The resulting difference between positive and negative $\lambda$ should be clear from Fig.~\ref{4PanelView}. For $\lambda>0$ we see a total of two bands that show up clearly. In the upper panel is a thin band that begins for small amplitude at $k=\sqrt{3}\,m$ and bends to the right; we shall explain this structure in Part 2 \cite{Part2}. In the lower panel we see a thick band that begins at small amplitude at $k=0$. This band continues to exists for small $k$ for any amplitude; we shall explain this structure in Section \ref{LongWavelength}.

For $\lambda<0$ we again see a total of two bands that show up clearly. In the upper panel is a thin band that again begins for small amplitude at $k=\sqrt{3}\,m$ and bends to the left; we shall explain this structure in Part 2 \cite{Part2}. In the upper panel we also see a thick band that begins at small amplitude at $k=0$.
These bands only make sense up to the maximum amplitude $\Phi_a$, but in this regime the band continues to exists for small $k$ for any amplitude; we shall explain this structure in Section \ref{LongWavelength}. Finally, in the lower panel, there is no additional instability.

\subsection{Adiabatic vs Isocurvature Behavior}

In the previous discussion we saw that there are two prominent instability bands; a rather thick band at small $k$ and a thin band that begins at $k=\sqrt{3}\,m$ (there should be even much thinner bands at higher $k$ also). This is true whether $\lambda$ is positive or negative. This gives the impression that positive or negative is qualitatively similar.

We would like to discuss that in fact there is a huge qualitative and quantitative difference between the positive and negative $\lambda$ cases. This is associated with the character of the modes that are being resonant. In particular, let us focus on the dominant thick band that extends towards $k=0$. This band is associated with $\delta\phi_\perp$ for $\lambda>0$ and $\delta\phi_\parallel$ for $\lambda<0$. These two classes of fluctuations are physically very different. In fact, as we will discuss in detail in Section \ref{LongWavelength}, the $\delta\phi_\parallel$ fluctuation is associated with an {\em adiabatic} mode, while the $\delta\phi_\perp$ fluctuations are associated with {\em isocurvature} modes. The adiabatic mode is characterized by a density perturbation, while the isocurvature mode is characterized by a conserved number density perturbation; we shall clarify all these details in Section \ref{LongWavelength}. Hence the sign of $\lambda$ determines whether it is the adiabatic or isocurvature modes that are resonant for long wavelengths. In this paper we shall get to the bottom of this interesting observation. 

As a consequence of these numerical results, it follows that in this case of a single field with $\lambda>0$, there would be no isocurvature mode, and hence relatively little instability. This is the classic observation that pure $\sim\lambda\phi^4$ inflation leads to inefficient resonance, as mentioned in the introduction. On the other hand, when there are multiple fields driving inflation, there will exist isocurvature modes, and hence there can be significant self-resonance even in classic models with $\lambda>0$.

\subsection{Negative Mass Squared Behavior}

\begin{figure}[t]
  \includegraphics[width=\cw]{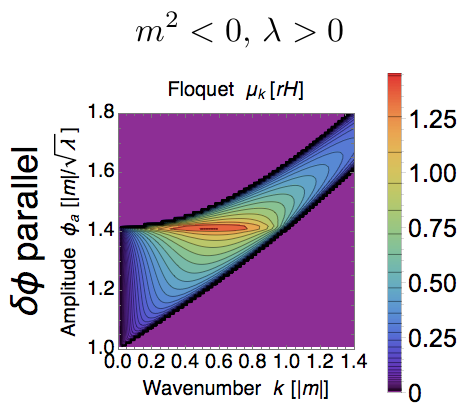}\\\vspace{0.2cm}
    \includegraphics[width=\cw]{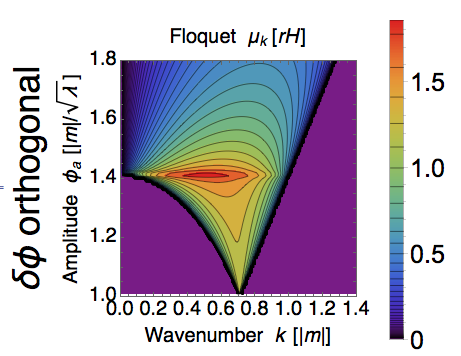}
  \caption{Contour plot of the real part of Floquet exponent $\mu_k$ for dimension 4 potentials as a function of wavenumber $k$ and background amplitude $\phi_a$ with $m^2<0$ and $\lambda>0$. Upper panel is for $\delta\phi_\parallel$ and lower panel is for $\delta\phi_{\perp}$. 
 We have plotted $\mu_k$ in units of $\Pv H$ where $\Pv\equiv \sqrt{\lambda}\,\mpl/|m|$, $k$ in units of $|m|$, and $\phi_a$ in units of  $|m|/\sqrt{\lambda}$.} 
\label{NegativeMass}\end{figure}

For $\lambda>0$ another possibility arises by allowing a tachyonic mass $m^2<0$. This gives rise to a type of Higgs potential. Here we choose $V_0=m^4/(4\lambda)$ in order to bring the energy density at the true vacuum to zero.

The numerical results for the Floquet exponent in this case is given in Fig.~\ref{NegativeMass}. Here we have chose the dimensionless amplitude $\Phi_a$ to be in the domain $\Phi_a\geq1$, i..e., the field amplitude can be taken to be greater than or equal to the field's vacuum expectation value $\phi_{vev}=|m|/\sqrt{\lambda}$.

We see that for the adiabatic mode $\delta\phi_\parallel$, the instability now begins at small amplitude at $k=0$, and for the isocurvature mode $\delta\phi_\perp$, the instability at small amplitude now begins at $k=|m|/\sqrt{2}$; as we will explain in Part 2 \cite{Part2}.

Furthermore, we see the complementary behavior that for small $k$, there is either stability/instability above the point $\phi_a = \sqrt{2}\,|m|/\sqrt{\lambda}$. This is the critical amplitude beyond which the field oscillates across the full double well potential. While for $|m|/\sqrt{\lambda}<\phi_a<\sqrt{2}\,|m|/\sqrt{\lambda}$ the background only oscillates on one side of the double well. This alters the effective sign of a type of pressure associated with the background; we shall discuss these sorts of matters and define the pressure in Section \ref{LongWavelength}.

\section{Auxiliary Potential for Isocurvature Modes} \label{AuxiliaryPotential}

In the previous section we showed numerical evidence that there is a significant difference between the behavior of the adiabatic modes and the isocurvature modes. This is especially true with regards to the existence or non-existence of a large instability band at long wavelengths. 

In the next section we will show how the behavior at long wavelengths of the adiabatic mode can be derived from a sound speed associated with the pressure and density of the background. Since this analysis will be so physical and intuitive, we would like to be able to discuss the isocurvature modes in a similar way. It is therefore important to be able to re-organize the equations that govern the isocurvature modes into a form that resembles those of the adiabatic modes. This will require the construction of a type of {\em auxiliary potential} for the isocurvature modes that we now describe.

\subsection{Correspondence between Modes} \label{Correspondence}

Recall the forms of the Hill's functions $h$ from eq.~(\ref{HillFunction}). We see that the $h$ driving the adiabatic mode $\delta\phi_\parallel$ is related to the $h$ driving the isocurvature mode $\delta\phi_\perp$ by the replacement
\beq
V''(\phi_0) \to {V'(\phi_0)\over\phi_0}
\eeq
We would like to bring the second expression into the same form as the first. To do so we need to construct an auxiliary potential $\tilde V$, with background solution $\phip_0$, for the isocurvature mode that satisfies
\beq
\tilde V''(\phip_0) = {V'(\phi_0)\over\phi_0}
\label{VpDef}\eeq
It is important to note that the primes here refer to each potential's respective arguments.

The equation of motion for $\phip_0$ is, by definition, the standard equation of motion with respect to its potential $\Vp$ (again ignoring Hubble)
\beq
\ddot\phip_0 + \Vp'(\phip) = 0
\eeq
Lets take a time derivative of this equation and use the chain rule
\beq
\dddot\phip_0 + \Vp''(\phip)\dot\phip = 0
\eeq
Then substituting eq.~(\ref{VpDef}) into this, we see that this corresponds to the equation of motion for $\phi_0$ if we identify $\phip_0$ as being related to $\phi_0$ in the following way
\beq
\dot\phip_0 = {\phi_0\over\Tp}
\label{phipSoln2}\eeq
where $\Tp$ is an arbitrary (non-zero) constant with units of time, whose value can be selected by convenience. Equivalently, this relationship can be solved for $\phip_0$ as
\beq
\phip_0(t)={1\over\Tp}\int^t dt'\,\phi_0(t')
\label{phipSoln1}\eeq

Another way to see this relationship between the pump $\phi_0(t)$ that controls the adiabatic mode and the pump $\phip_0(t)$ that controls the isocurvature mode is as follows: In Section \ref{LongWavelength} we will relate $\delta\phi_\parallel$ to the energy density perturbation $\delta\ed$, and relate $\delta\phi_\perp$ to the number density perturbation $\delta\nd$; these definitions and relationships shall be discussed there. We find that (again ignoring Hubble expansion for now) the linearized equations of motion for these perturbations are
\bea
&&\ddot{\delta\ed}-2{\ddot\phi_0\over\dot\phi_0}\dot{\delta\ed}+k^2\delta\ed = 0\label{edFric}\\
&&\ddot{\delta\nd}_i-2{\dot\phi_0\over\phi_0}\dot{\delta\nd}_i+k^2\delta\nd_i = 0\label{ndFric}
\eea
where we are again in $k$-space. So we see quite directly that to pass from $\delta\ed$ to $\delta\nd$ requires replacing $\dot\phi_0$ by $\phi_0$ (up to a multiplicative constant) in agreement with eq.~(\ref{phipSoln2}). In fact one can go further and construct a quadratic action for each of these physical variables of the form
\beq
S[\delta] = \int d^4 x {1\over f^2(t)}\left[{1\over 2}\dot\delta^2-{1\over2}(\nabla\delta)^2 \right]
\eeq
where $f(t)\propto\dot\phi_0(t)$ for $\delta\to\delta\ed$ and $f(t)\propto\phi_0(t)$ for $\delta\to\delta\nd$, again showing the correspondence.

\subsection{Integral Form for General Potentials}

We now show how to solve for the auxiliary potential $\Vp$ for any potential $V$. For simplicity, we assume that the true minimum of the potential is at $\phi=0$. However, an extension to the tachyonic mass $m^2<0$ cases is straightforward.

Firstly, since the potential $V$ is assumed to carry an internal rotational symmetry, it should be some series in $\phi_0^2$, rather than having any odd powers of $\phi_0$. To make this explicit, it is useful to introduce the variable
\beq
\phis_0\equiv{1\over2}\phi_0^2
\eeq
where the factor of 1/2 is for convenience. Using the chain rule, eq.~(\ref{VpDef}) may be rewritten as
\beq
\tilde V''(\phip_0) -{\partial V\over\partial\phis_0} = 0
\label{Vpeqn}\eeq
Using the relationship (\ref{phipSoln2}) and the conservation of energy of the $\phip_0$ field, we can rewrite $\phis_0$ as
\beq
\phis_0 = {1\over 2}\Tp^2\dot\phip^2 = \Tp^2(\Vp(\phip_a)-\Vp(\phip_0))
\eeq
where we have introduced the amplitude of the $\phip_0$ oscillations as $\phip_a$. Using the chain rule, we can then rewrite (\ref{Vpeqn}) as
\beq
\tilde V''(\phip_0) +{1\over \Tp^2}{\partial V\over\partial\Vp} = 0
\eeq
Now this has the structure of an equation of motion for $\Vp$ as a function of $\phip$ driven by a potential $V/\Tp^2$. Such an equation always possesses a first integral, which is
\bea
{1\over2}\Vp'(\phip_0)^2+{1\over\Tp^2}V\!\left(\Tp^2(\Vp(\phip_a)-\Vp(\phip_0))\right)={1\over2}\Vp'(\phip_a)^2\,\,\,\,\,\,\,\,\,\,\,
\label{1stIntegral}\eea
Now using the equation of motion evaluated at $\phip_0=\phip_a$ (where $\dot\phip_0=0$) we obtain
\beq
{1\over2}\Vp'(\phip_a)^2 = {1\over\Tp^2}V(\phis_a)
\eeq
and the relationship between the amplitudes is
\beq
\Vp(\phip_a) = {\phis_a\over\Tp^2}
\eeq
Inserting this into (\ref{1stIntegral}) allows us to construct the following integral solution
\beq
\int_0^{\Vp}{d\vp\over\sqrt{2V(\phis_a)-2V(\phis_a-\Tp^2\vp)}} = {\phip_0\over\Tp}
\label{IntegralSoln}\eeq
Note that in the integrand, the symbol $\vp$ is the ``dummy variable" of integration. In principle, for a given choice of $V$ and amplitude $\phi_a$, this integral can be performed and inverted to find the auxiliary potential $\Vp=\Vp(\phip_0)$. It is important to note that such a potential will depend on the choice of amplitude $\phi_a$.

\subsection{Application to Dim 4 Potentials}\label{AppDim4}

Let us illustrate this with the dimension 4 potentials we analyzed in Section \ref{Quartic}. Recall that (for $m^2>0$) the potential is 
\beq
V(\phi)={1\over2}m^2\phi^2+{1\over4}\lambda\,\phi^4
\label{V4}\eeq
When rewritten in terms of the $\phis$ variable, this is $V(\phis) = m^2\phis+\lambda\,\phis^2$.
We substitute this into the integral solution of eq.~(\ref{IntegralSoln}) and carry out the integral. We find the integral is an inverse cosine. Upon inversion, the resulting auxiliary potential for the isocurvature modes is
\beq
\Vp(\phip) = {m^2+\lambda\,\phi_a^2\over 2\,\lambda\,\Tp^2}\left(1-\cos(\sqrt{2\lambda}\,\Tp\,\phip)\right)
\label{VA4}\eeq
This representation is useful for $\lambda>0$. While for $\lambda<0$ we can rewrite it as
\beq
\Vp(\phip) = {m^2-|\lambda|\phi_a^2\over 2|\lambda|\,\Tp^2}\left(\cosh(\sqrt{2\lambda}\,\Tp\,\phip)-1\right)
\eeq
Also the auxiliary field amplitude $\phip_a$ can be determined from the original physical field's amplitude $\phi_a$ by
\beq
\phip_a = {1\over\sqrt{2\lambda}\,\Tp}\tan^{-1}\!\left(\sqrt{{2\lambda\phi_a^2\over m^2}+{\lambda^2\phi_a^4\over m^4}}\right)
\eeq
When $\lambda<0$ this becomes an inverse hyperbolic tangent function.

\begin{figure*}[t]
  \includegraphics[width=\cw]{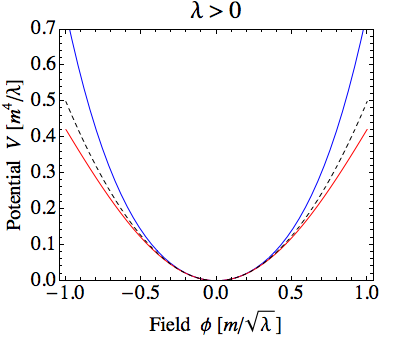}
    \includegraphics[width=\cw]{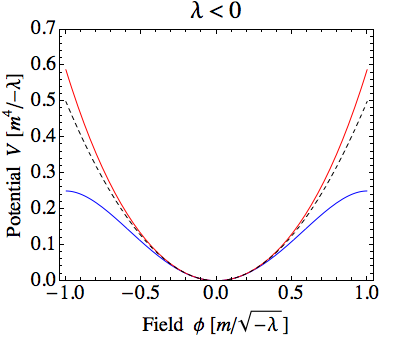}
  \caption{Potential function for the dimension 4 theory with $m^2>0$. Left panel is $\lambda>0$ and right panel is $\lambda<0$. The blue curves are the fundamental potentials $V(\phi)$, the red curves are the auxiliary potentials $\Vp(\phip)$, and the black dashed curves are the quadratic potentials $\sim{1\over2}m^2\phi^2$. We have taken $\Tp=1/m$ and fixed $\phi_a$ to be small to define the $\Vp$ function.}
\label{PotentialPlot}\end{figure*}

A plot of the original potential $V$ that controls the adiabatic mode and the auxiliary potential $\Vp$ that controls the isocurvature modes is given in Fig.~\ref{PotentialPlot} for $\lambda>0$ (left panel) and $\lambda<0$ (right panel). We see the complementary behavior of the potentials. Compared to a quadratic potential $\sim{1\over2}m^2\phi^2$, for $\lambda>0$, $V$ grows more quickly and $\Vp$ grows more slowly, while for $\lambda<0$, $V$ grows more slowly and $\Vp$ grows more quickly.

For small amplitudes, $\phi_a\ll m/\!\sqrt{|\lambda|}$, we can Taylor expand the auxiliary potential $\Vp$. For convenience we pick $\Tp=1/m$ and we find
\beq
\Vp(\phip) = {1\over 2}m^2\phip^2-{1\over12}\lambda\,\phip^4+\ldots
\label{V4A}\eeq
Comparing eq.~(\ref{V4}) to eq.~(\ref{V4A}) we see that the quartic coupling $\lambda$ has been effectively replaced by 
\beq
\lambda\to-{\lambda\over3}
\eeq
We shall also see derive this result from a small amplitude analysis in Part 2 \cite{Part2}.

\section{General Analysis for Long Wavelength Modes} \label{LongWavelength}

In this Section we show that the shape of the above potentials ($V$ and $\Vp$) rather directly determines the existence or non-existence of a large instability band at long wavelengths. To do so we first discuss the physical structure of the modes and then prove general results about their behavior based on pressure and density arguments.

\subsection{Goldstone Modes}

The oscillating background $\vec\phi_0(t)$ breaks two important symmetries of the underlying theory. Firstly, since it is time dependent, it breaks time translation symmetry. Secondly, since its motion is radial, it must choose some direction in field space and hence it breaks the internal rotational symmetry. This has important consequences at long wavelengths. The low energy states of the theory are subject to the {\em Goldstone theorem}, which requires that each of these broken symmetries is associated with {\em massless modes}. 

Note that the theorem applies to the low energy, or ``effective" theory, which is applicable at long wavelengths. In order to construct the effective theory, we shall have to perform a type of time averaging in order to coarse grain the system sufficiently; we shall see this in the upcoming Sections \ref{AdiMode}, \ref{IsoMode}.

The associated massless (or ``gapless") modes are associated with the corresponding conserved quantities. In particular, the Goldstone mode associated with the breaking of time translation symmetry is the {\em energy density} $\delta\ed$, since the integrated energy density is conserved; this is an adiabatic mode. While the Goldstone modes associated with the breaking of the internal rotational symmetry are the {\em number densities} $\delta\nd_i$, since the integrated number densities are conserved; these are isocurvature modes. We shall rigorously count the number of these isocurvature modes and construct all these various quantities carefully in the next subsections.

\subsection{Adiabatic Mode $\delta\ed$} \label{AdiMode}

Let us begin be constructing the full energy density stored in the field $\vec\phi$. It is given by
\beq
\ed = {1\over2}|\dot{\vec\phi}|^2+{1\over2}|\nabla\vec\phi|^2+V(\vec\phi)
\eeq
where we allow for $N$ fields and a potential $V$ that carries an internal symmetry, as before.
The homogeneous background energy density is given by
\beq
\ed_0 = {1\over2}\dot\phi_0^2+V(\phi_0)
\eeq
where the field $\vec\phi_0$ is assumed to point in a specific direction, such as eq.~(\ref{BgdDirection}). 

The first order perturbation is given by
\beq
\delta\ed = \left(\dot\phi_0{\partial\over\partial t}+V'(\phi_0)\right)\delta\phi_\parallel
\label{ed1st}\eeq
We see that the energy density is some linear time dependent operator acting on the parallel perturbation $\delta\phi_\parallel$ and is independent of the orthogonal perturbations $\delta\phi_{\perp i}$ at this order. Instead we shall see that $\delta\phi_{\perp i}$ is relevant for the modes of Section \ref{IsoMode}. Hence perturbations in $\delta\phi_\parallel$ cause energy density perturbations without affecting the relevant abundance of particle species. So this is, by definition, an {\em adiabatic} mode.

\subsubsection{Equation of Motion}

We would like to construct a second order equation of motion for the energy density perturbation $\delta\ed$.
To do so we take time derivatives of the expression for $\delta\ed$ in eq.~(\ref{ed1st}). The first time derivative can be written as 
\beq
\dot{\delta\ed} = \left(\ddot\phi_0+V'(\phi_0)\right)\dot{\delta\phi}_\parallel+\left(\ddot{\delta\phi}_\parallel+V''(\phi_0)\delta\phi_\parallel\right)\dot\phi_0
\eeq
The first term in parenthesis vanishes by the equation of motion for $\phi_0$, while the second term in parenthesis can be simplified by the equation of motion for $\delta\phi_\parallel$. This gives
\beq
\dot{\delta\ed} = - k^2\dot\phi_0\delta\phi_\parallel
\label{ed2nd}\eeq
We now take another time derivative giving
\beq
\ddot{\delta\ed} = - k^2\ddot\phi_0\delta\phi_\parallel-k^2\dot\phi_0\dot{\delta\phi}_\parallel
\eeq
We now use eq.~(\ref{ed1st}) to eliminate $\dot\phi_0\dot{\delta\phi}_\parallel$ and the equation of motion for $\phi_0$ to eliminate $\ddot\phi_0$, giving
\beq
\ddot{\delta\ed} + k^2(\delta\ed- 2 V'(\phi_0)\delta\phi_\parallel) = 0
\label{ed3rd}\eeq
This now begins to take on the form of a wave equation for $\delta\ed$ (recall $k^2\to -\nabla^2$) for a massless mode, however the term $V'(\phi_0)\delta\phi_\parallel$ prevents this from being precise. One way to proceed, is to now eliminate $\delta\phi_\parallel$ using eq.~(\ref{ed2nd}). This leads to eq.~(\ref{edFric}) that we mentioned earlier. However, in order to organize this properly into a wave equation, we prefer to keep this form of the second order equation for $\delta\ed$ and proceed to do some form of coarse graining, as we now describe. 

\subsubsection{Time Average} \label{TimeAdi}

In this context, coarse graining means to average over sufficiently long time scales; this loses information for high $k$ (high frequency modes) but allows us to probe the long wavelength physics. To do this we perform a time average of eq.~(\ref{ed3rd}) over one period of the background. 

Now, in the infinite wavelength limit, $\delta\ed$ becomes uniform in space and therefore it must be constant in time since energy is conserved. This means that for sufficiently long wavelengths, $\delta\ed$ should be {\em slowly varying} in time compared to $\delta\phi_\parallel$; the energy density is plotted later in Fig.~\ref{Growth} where we see its slow evolution in the upper right panel, compared to the rapid oscillations in the parallel fluctuations in the upper left panel. Hence when we average over the rapid oscillation of the background $\phi_0(t)$ there will be negligible alteration in $\delta\ed$, i.e.,
\beq
\langle\delta\ed\rangle \approx \delta\ed
\eeq
On the other hand, we have to be very careful when we time average the term $V'(\phi_0)\delta\phi_\parallel$ in eq.~(\ref{ed3rd}), since both $V'(\phi_0)$ and $\delta\phi_\parallel$ are rapidly varying in time; ahead in Fig.~\ref{Growth} we plot this rapid of oscillation of $\delta\phi_\parallel$ in the upper left panel. 
So we have
\beq
\ddot{\delta\ed} + k^2(\delta\ed- 2 \langle V'(\phi_0)\delta\phi_\parallel\rangle) = 0
\label{ed3rdTA}\eeq
Now the term that we require to time average is multiplied by $k^2$. So at long wavelengths, we may evaluate this quantity in the $k\to0$ limit, for otherwise we would be tracking sub-leading corrections. In this limit, such a quantity can only be a function of the amplitude that we let $\phi$ fall from; this is a combination of the background amplitude $\phi_a$ and a perturbation. Similarly, the energy density itself is only a function of the amplitude in this long wavelength limit. Hence we must be able to trade one for the other. Using a type of ``chain rule" this is
\beq
\langle V'(\phi_0)\delta\phi_\parallel\rangle = {d\langle V\rangle\over d\langle\ed_0\rangle}\delta\ed
\eeq
where $\langle V\rangle$ and $\langle \ed_0\rangle$ are the time average of the potential and energy density evaluated on the background solution $\phi_0$, respectively. Substituting this into (\ref{ed3rdTA}) gives
\beq
\ddot{\delta\ed} + k^2\left(1- 2 {d\langle V\rangle\over d\langle\ed_0\rangle}\right)\delta\ed = 0
\label{ed3rdTA2}\eeq
which is indeed of the form of a wave equation. Note that by construction, the time averaged quantities in brackets here are time and space independent.

\subsubsection{Sound Speed $c_S$} \label{SoundSpeed}

Now it is useful to express the above derivative in terms of a more physical quantity; the time averaged {\em pressure}. The time averaged pressure and energy density of the background are given by
\bea
&&\langle \pr_0 \rangle = \left\langle{1\over 2} \dot\phi_0^2\right\rangle - \langle V \rangle\\
&&\langle \ed_0 \rangle = \left\langle{1\over 2} \dot\phi_0^2\right\rangle + \langle V \rangle
\eea
So the difference is 
\beq
\langle \pr_0 \rangle - \langle\ed_0\rangle = - 2\langle V\rangle
\label{p1}\eeq
A derivative with respect to $\langle\ed_0\rangle$ evidently gives
\beq
c_S^2 = 1- 2 {d\langle V\rangle\over d\langle\ed_0\rangle} 
\eeq
where
\beq
c_S^2 \equiv {d\langle \pr_0 \rangle\over d\langle\ed_0\rangle}
\label{cs}\eeq
is the sound speed squared.
Substitution into eq.~(\ref{ed3rdTA2}) leads to the sound wave equation
\beq
\ddot{\delta\ed} + c_S^2\,k^2\,\delta\ed = 0
\eeq
This proves that indeed the adiabatic perturbations have a gapless spectrum, even though the field fluctuations $\delta\phi_\parallel$ generally do not.

This shows that stability or instability of $\delta\ed$ is determined by the value of the squared sound speed $c_S^2$. If $c_S^2>0$, then long wavelengths modes will oscillate. On the other hand, if $c_S^2<0$, then long wavelengths modes will grow exponentially.
Indeed we can identify the Floquet exponents as
\beq
\mu_k = \pm \, i \,c_S\,k
\eeq
We see that the strength of the instability vanishes in the $k\to0$ limit, but this band can still be very important at small, but finite $k$, as we saw numerically in the previous section.

Now in order to evaluate $c_S^2$ we need a recipe to evaluate $\langle\ed_0\rangle$ and $\langle \pr_0 \rangle$. It is useful to express these as functions of the amplitude of oscillation $\phi_a$. For the energy density $\langle\ed_0\rangle$ it is trivial because energy is conserved, giving
\beq
\langle\ed_0\rangle = V(\phi_a)
\eeq
For the pressure $\langle \pr_0 \rangle$ it is more non-trivial since pressure oscillates throughout the background cycle. Using (\ref{p1}) we may write it as 
\beq
\langle \pr_0 \rangle = V(\phi_a) - {2\over T(\phi_a)}\int_0^{T(\phi_a)} \!\!\!dt\, V(\phi_0(t))
\label{pressInt}\eeq
where $T$ is the period of the pump and $\int_0^T dt\,V$ is the integrated potential over a cycle. Using the equations of motion, they can be expressed as
\bea
T(\phi_a) &=& \int_{\phi_b}^{\phi_a}{d\phi_0\sqrt{2}\over\sqrt{V(\phi_a)-V(\phi_0)}}\label{TInt}\\
\int_0^{T(\phi_a)} \!\!\!dt\, V(\phi_0(t)) &=& \int_{\phi_b}^{\phi_a}{d\phi_0\sqrt{2}\,V(\phi_0)\over\sqrt{V(\phi_a)-V(\phi_0)}}\,\,\,\,\,\,\,
\label{VInt}\eea
where $\phi_b$ is the amplitude the field reaches on the other side of its potential. For most applications, we will consider expanding around a symmetric point, giving $\phi_b = -\phi_a$, but if we consider the $m^2<0$ case, then the relationship is more complicated.

Then with the (time averaged) pressure and energy density given as function of amplitude $\phi_a$ the sound speed square can be computed using the chain rule
\beq
c_S^2 = {d\langle \pr_0 \rangle\over d\phi_a} \cdot \left({d\langle \ed_0 \rangle\over d\phi_a}\right)^{\!-1}
\label{csChain}\eeq
Since energy density is taken to be an increasing function of amplitude, the sign of $c_S^2$ is determined by the sign of the derivative of pressure with respect to amplitude. This leads to a very physical understanding of the fate of the adiabatic mode: If pressure increases with amplitude, the mode is stable. If pressure decreases with amplitude, the mode is unstable. Furthermore, since the vacuum energy is taken to be zero, then the pressure is zero for zero amplitude. Hence, for small amplitudes, this can be expressed even more simply as: positive pressure implies stability and negative pressure implies instability.

\subsubsection{Application to Dim 4 Potentials}

Let us apply this formalism to the dimension 4 potentials of Section \ref{Quartic}. For now we consider $m^2>0$, the vacuum energy $V_0=0$, and allow $\lambda$ to be either positive or negative. The potential is then given by eq.~(\ref{V4}).
The sign of $\lambda$ should determine stability as it determines the sign of the pressure.

We are able to express the above integrals for $\langle \pr_0 \rangle$ in eq.~(\ref{pressInt}) in terms of elliptic integrals.
After doing so, we find the following result for the sound speed as a function of amplitude
\bea
c_S^2 = {(2m^2+\lambda\phi_a^2)[m^2 E(\psib_a)-(m^2+\lambda\phi_a^2)K(\psib_a)]^2\over 3\lambda\phi_a^2(m^2+\lambda\phi_a^2)^2 K(\psib_a)^2}\,\,\,\,\,\,\,\,\,\,\,
\label{csElliptic}\eea
where 
\beq
\psib_a \equiv -{\lambda\phi_a^2\over 2m^2+\lambda\phi_a^2}
\eeq
and $K$ and $E$ are the complete elliptic integrals of the first and second kind, respectively.
Note all the various squared factors in eq.~(\ref{csElliptic}). This means that the sign of $c_S^2$ is determined by the sign of $\lambda$. So we see explicitly that the sign of $\lambda$ determines the sign of the pressure.
For $\lambda>0$ and $\phi_a\gg m/\sqrt{\lambda}$ this expression collapses to $c_S^2=1/3$; we shall return to this in the next subsubsection. 

\begin{figure*}[t]
  \includegraphics[width=\cw]{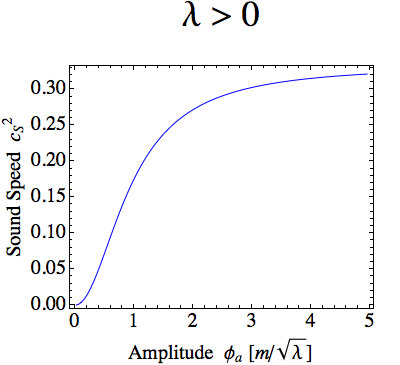}\,
   \includegraphics[width=\cw]{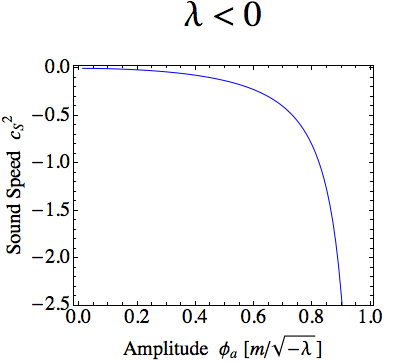}\vspace{0.2cm}
   \includegraphics[width=\cw,height=\htt]{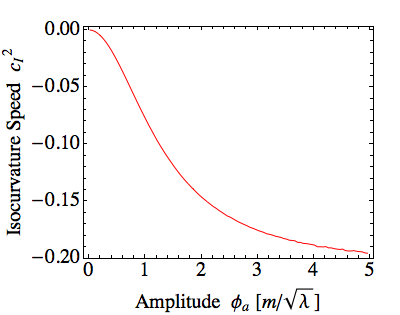}\,
   \includegraphics[width=0.98\cw]{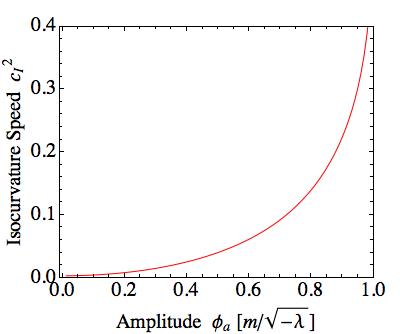}
\caption{The squared speeds as a function of amplitude $\phi_a$ in units of $m/\sqrt{|\lambda|}$ for the dimension 4 theory with $m^2>0$. 
Left panel is $\lambda>0$ and right panel is $\lambda<0$. Upper panel is the sound speed $c_S^2$ governing the stability of the adiabatic mode $\delta\ed$ ($\delta\phi_\parallel$). Lower panel is the speed $c_I^2$ governing the stability of isocurvature modes $\delta\nd_i$ ($\delta\phi_{\perp i}$).}
\label{SoundSpeedPlot}\end{figure*}

For general amplitudes, we plot $c_S^2$ in the upper panel of Fig.~\ref{SoundSpeedPlot}. We see that the sound speed (and hence the Floquet exponent) begins at zero for zero amplitude. This makes sense, because for small field amplitudes, the theory is approximately matter dominated, which has zero pressure. On the other hand, $c_S^2$ moves away from zero at finite amplitude. For $\lambda<0$ there is a corresponding instability due to the negative pressure, which becomes arbitrarily large near the hilltop $\phi_a\to\phi_{a,max}$. In general we expect there to be higher order corrections to the potential to provide a sensible model for inflation; this will weaken the strength of this instability. 

By recalling $\mu_k = \pm\,i\,c_S\,k$, this result for $c_S^2$ adequately explains the presence of the thick instability band we saw earlier in Fig.~\ref{4PanelView} for $\delta\phi_\parallel$ and $\lambda<0$.

\subsubsection{Application to Power Law Potentials}\label{PurePowerAdi}

\begin{figure}[t]
  \includegraphics[width=\cw]{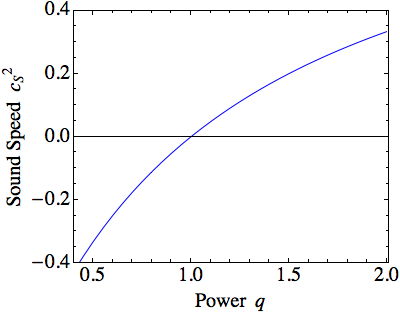}\\\bigskip
    \includegraphics[width=\cw]{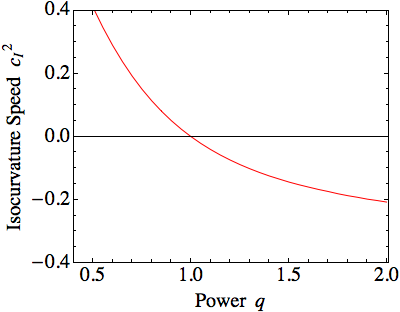}
\caption{The squared speeds as a function of the power $q$ for a pure power law potential.
The upper panel is the sound speed $c_S^2$ governing the stability of the adiabatic mode $\delta\ed$ ($\delta\phi_\parallel$). The lower panel is the speed $c_I^2$ governing the stability of isocurvature modes $\delta\nd_i$ ($\delta\phi_{\perp i}$).}
\label{SpeedsPowerLaw}\end{figure}

Let us now consider the case of a pure power law potential
\beq
V(\vec\phi) = {\lamh\over 2\,\ex}|\vec\phi|^{2\ex}
\label{Vpower}\eeq
with $\lamh>0$.
For this to involve ordinary operators around $\phi=0$, we expect $\ex$ to be an integer. However, we can also imagine that this power law is only the behavior of the potential at large field values, so we might allow $\ex$ to be any positive number. Indeed the coupling $\lamh$ may not be the same as the leading interaction coupling $\lambda$ from expanding around small field values. In any case, the integrals (\ref{TInt},\,\ref{VInt}) can be done analytically and the result yields
\beq
c_S^2 = {\ex-1\over \ex+1}
\eeq
A plot of $c_S^2$ for the power law potential is given in the upper panel of Fig.~\ref{SpeedsPowerLaw}.
For $\ex\geq1$, $c_S^2\geq0$, and we have stability. For example, for the quartic theory $\ex=2$, $c_S^2=1/3$, as is appropriate for a radiation era.
On the other hand, for $0<\ex<1$ we have an instability. This would be relevant to some models of inflation, such as ``axion monodromy models" \cite{McAllister:2008hb,ModMcAllister} where possible values of the power include $\ex=1/2,\,1/3$. 

Now assuming $0<\ex<1$, we have a non-zero and real Floquet exponent $\mu_k$ whose value is independent of amplitude in this small $k$ approximation (since $c_S^2$ is a constant). But we need to know the ratio of $\mu_k$ to the Hubble parameter $H$. For the power law potential (\ref{Vpower}), we have $H\sim \sqrt{\lamh}\,\phi_a^\ex/\mpl$, giving $\mu_k/H\sim \mpl k/(\sqrt{\lamh}\,\phi_a^\ex$). This ratio becomes arbitrarily large at small field amplitudes. However, at some point a realistic potential should transition from this fractional power law to say a regular quadratic potential at small field values. Lets call the transition scale $\phi_a\sim \F$; which acts as a cutoff on the field theory. A toy example of this behavior is \cite{Amin:2011hj}
\beq
V(\phi) = {m^2 \F^2\over 2\,\ex}\left(\left({|\vec\phi|^2\over \F^2}+1\right)^{\!\ex}-1\right)
\label{toy}\eeq
This implies that $\lamh$ will be related to the transition scale $\F$ and mass $m$ by $\lamh\sim m^2/\F^{2(\ex-1)}$.
At the transition regime, we obtain $\mu_k/H\sim k\,\mpl/(m\,\F)$. Now the dominant instability occurs when $k$ is ``small", but parametrically of the same order as $m$; see Fig.~\ref{4PanelView}. So then we have $\mu_k/H\sim\mpl/\F$. Hence a large instability is associated with the transition scale satisfying $\mpl/\F\gg1$. In fact, by Taylor expanding around small $\phi$, we can relate $\F$ to the quartic coupling $\lambda$ by $\F\sim m/\sqrt{|\lambda|}$.
So to use the notation of Section \ref{Quartic}, this corresponds to the statement that $\Pv\equiv\sqrt{|\lambda|}\,\mpl/m\gg 1$ for large instability.

\subsection{Isocurvature Modes $\delta\nd$} \label{IsoMode}

In the previous section we studied the energy density; the density associated with the conserved energy. In this section we study the various number densities; the densities associated with conserved particle numbers.

The $O(\N)$ internal symmetry $\phi^i\to R^i_j\phi^j$ leads, by the Noether theorem, to the following set of number densities
\beq
\nd_{ij} = \dot\phi_i\,\phi_j - \dot\phi_j\,\phi_i
\eeq
The integral over space $\Delta \Num_{ij}=\int d^3x\,\nd_{ij}$ is a set of $\N(\N-1)/2$ conserved particle numbers. For a complex field ($\N=2$) $\Delta\Num$ is the number of particles minus the number of antiparticles; this will be examined further in Part 2 \cite{Part2} with regards to its possible relation to baryogenesis.

Now let us expand around the background $\vec\phi_0$ given by eq.~(\ref{BgdDirection}). To leading non-zero order, we have the following set of $\N-1$ linear quantities
\beq
\delta\nd_{i} = -\left(\phi_0{\partial\over\partial t}-\dot\phi_0\right)\delta\phi_{\perp i}
\label{ni}\eeq
with $i=1,\ldots,\N-1$. Also, to leading non-zero order, we have the following set of $(\N-1)(\N-2)/2$ quadratic quantities
\beq
\delta\nd_{ij} = \!\dot{\,\delta\phi_{\perp i}}\,\delta\phi_{\perp j} - \!\dot{\,\delta\phi_{\perp j}}\, \delta\phi_{\perp i}
\eeq
with $i,j=1,\ldots,\N-1$. This latter set of conserved quantities will not appear in the leading order analysis of the low lying modes.
Instead the modes of interest are the $\N-1$ densities $\nd_i$. We see that these densities $\nd_i$ are given by some linear operator acting on the orthogonal perturbations $\delta\phi_{\perp i}$ and are independent of the parallel perturbations $\delta\phi_\parallel$ at this order. Hence perturbations in $\delta\phi_\perp$ cause perturbations in the relative number densities of species without affecting the total energy density. So these are, by definition, {\em isocurvature} modes.

\subsubsection{Equation of Motion}

As we did before for $\delta\rho$, we would like to construct a second order equation of motion for the number density perturbations $\delta\nd_i$. A first time derivative gives
\beq
\dot{\delta\nd}_i = \ddot\phi_0\,\delta\phi_{\perp i}-\phi_0 \ddot{\,\delta\phi_{\perp i}}
\label{nd2nd}\eeq
Then using the equation of motion for $\phi_0$ and the equation of motion for $\delta\phi_{\perp i}$, this can be simplified to
\beq
\dot{\delta\nd}_i = k^2\phi_0\,\delta\phi_{\perp i}
\eeq
We now take another time derivative and use eq.~(\ref{ni}) to eliminate $\!\dot{\,\delta\phi_{\perp i}}$. This gives the second order equation
\beq
\ddot{\delta\nd}_i+k^2(\delta\nd_i-2\dot\phi_0\,\delta\phi_{\perp i}) = 0
\eeq
This result is analogous to eq.~(\ref{ed3rd}) that we obtained for the energy density perturbation $\delta\ed$.
If we use eq.~(\ref{nd2nd}) to eliminate $\delta\phi_{\perp i}$ in favor of $\delta\nd_i$ we obtain the second order equation for $\delta\nd_i$ that we mentioned earlier in (\ref{ndFric}). But we would like to perform a time averaging of this present equation analogously to our time averaging of $\delta\ed$.

\subsubsection{Time Average with Auxiliary Potential} \label{TimeIso}

\begin{figure*}[t]
  \includegraphics[width=\cw,height=\htz]{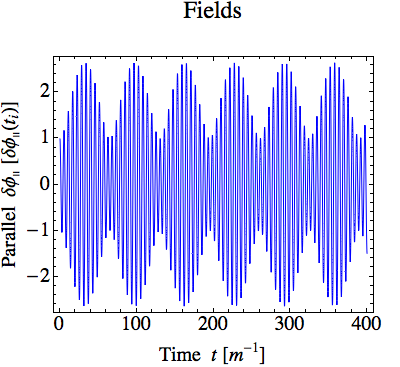}\,
   \includegraphics[width=\cw]{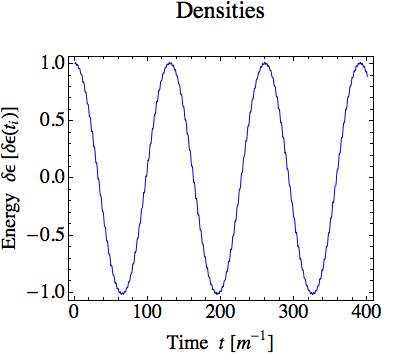}\\
   \includegraphics[width=\cw,height=\htt]{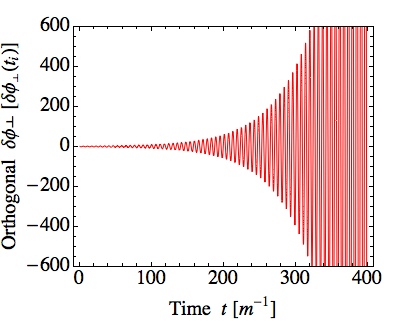}\,
   \includegraphics[width=\cw]{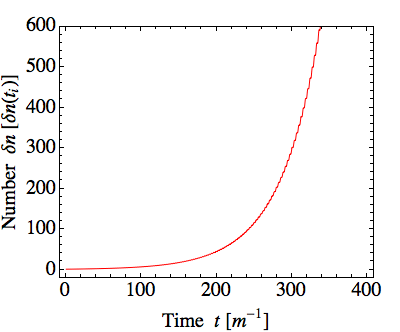}
\caption{Representative plot of the time evolution of the fluctuations in the dimension 4 potentials, with $m^2>0$ and $\lambda>0$.
Left panel are the fields (rapidly oscillating) and right panel are the densities (slowly varying). 
Upper panel is for the adiabatic mode: $\delta\phi_\parallel$ and $\delta\ed$.
Lower panel is for the isocurvature mode: $\delta\phi_\perp$ and $\delta\nd$.
We have plotted each fluctuation in units of its initial starting value and time in units of inverse mass. 
For definiteness, we chose a background amplitude of $\phi_a = 0.4\,m/\sqrt{\lambda}$ and wavenumber $k=0.2\,m$; this is in the regime of stability for the adiabatic mode and instability for the isocurvature mode (see left panel of Fig.~\ref{4PanelView}). For $\lambda<0$ (not shown here) the stability structure is interchanged.}
\label{Growth}\end{figure*}

As earlier, in order to make progress, we consider long wavelengths. If we went to infinite wavelengths, then $\delta\nd_i$ would be constant since number densities are conserved by the Noether theorem. So for sufficiently long wavelengths, the number density should be {\em slowly varying} in time compared to $\delta\phi_{\perp i}$; the number density is plotted in Fig.~\ref{Growth} where we see its relatively slow variation in the lower right panel, compared to the rapid oscillation in the orthogonal field fluctuations in the lower left panel. It is true that (for the parameters chosen) $\delta\nd_i$ is growing exponentially, but the growth rate is small compared to the period of $\delta\phi_{\perp i}$ in this long wavelength regime.
So if we time average over the period of background oscillation we have
\beq
\langle\delta\nd_i\rangle\approx\delta\nd_i
\eeq
We are then led to the time averaged equation
\beq
\ddot{\delta\nd}_i+k^2(\delta\nd_i-2\langle\dot\phi_0\,\delta\phi_{\perp i}\rangle) = 0
\label{ndTA}\eeq
where we must deal with rapidly oscillating factors $\dot\phi_0$ and $\delta\phi_{\perp i}$; this rapid oscillation is seen in the lower left panel of Fig.~\ref{Growth}. 

In this form it appears difficult to re-organize this into a useful form. So this is where we make use of the correspondence between $\delta\nd_i$ and $\delta\ed$. The idea is to re-express this in terms of the auxiliary field $\phip_0$ variable we introduced in Section \ref{Correspondence}. Then treat the isocurvature fluctuation with respect to the potential $V$ as analogous to an adiabatic perturbation with respect to the auxiliary potential $\Vp$. Now let's recall the relationship $\dot\phip_0=\phi_0/\Tp$, which implies
\beq
\dot\phi_0 = \Tp\,\ddot\phip_0 = -\Tp\,\Vp'(\phip_0)
\eeq
Substituting into (\ref{ndTA}) then gives
\beq
\ddot{\delta\nd}_i+k^2(\delta\nd_i+2\,\Tp\,\langle\Vp'(\phip_0)\delta\phi_{\perp i}\rangle) = 0
\label{ndTA2}\eeq
This now has a similar structure to eq.~(\ref{ed3rdTA}) for the time averaged $\delta\ed$. So we can again make use of a type of chain rule to express the final time averaged quantity in terms of the auxiliary background quantities, namely
\beq
-\Tp\,\langle \Vp'(\phip_0)\delta\phi_{\perp i}\rangle = {d\langle \Vp\rangle\over d\langle\edp_0\rangle}\delta\nd_i
\eeq
Note the negative sign here is because of the relative overall sign between the expression for the linearized densities in (\ref{ed1st}) and (\ref{ni}).
Upon substitution we have
\beq
\ddot{\delta\nd}_i+k^2\left(1- 2 {d\langle \Vp\rangle\over d\langle\edp_0\rangle}\right)\delta\nd_i = 0
\label{nd3rdTA2}\eeq
Hence we are led to a direct correspondence between the wave equation for $\delta\ed$ of (\ref{ed3rdTA2}) and a wave equation for $\delta\nd_i$.

\subsubsection{Isocurvature Speed $c_I$}

From this wave equation, we can identify a speed from the auxiliary pressure $\pp_0$ and energy density $\edp_0$. This analysis goes through as in Section \ref{SoundSpeed}, so we do not repeat all the details here. It suffices to say that there is a type of isocurvature speed given by
\beq
c_I^2 \equiv {d\langle\pp_0\rangle\over d\langle\edp_0\rangle}
\label{ci}\eeq
with the wave equation given by
\beq
\ddot{\delta\nd}_i+c_I^2\, k^2 \,\delta\nd_i = 0
\label{ndW}\eeq
So stability of the isocurvature modes at long wavelengths is determined by the sign of $c_I^2$, with the corresponding Floquet exponent given by
\beq
\mu_k = \pm \, i\, c_I\, k
\eeq

Now there is an important technical difference between the way we need to compute the derivative of the auxiliary pressure in eq.~(\ref{ci}) compared to how we computed the derivative of pressure in eq.~(\ref{cs}). For the adiabatic mode, we previously made use of the chain rule to rewrite the derivatives with respect to the pump amplitude $\phi_a$ of interest in eq.~(\ref{csChain}). However, for the isocurvature modes, we need to be careful since the auxiliary potential $\Vp$, and hence the auxiliary pressure $\pp_0$ and energy density $\edp_0$, depend on both the field $\phip$ and the amplitude itself; even before time averaging. In order to define a physical derivative we need to fix the {\em theory} as we vary the {\em amplitude}. To make the dependence on amplitude explicit, lets write the auxiliary potential as
\beq
\Vp = \Vp(\phip,\phi_a^*)
\eeq
where the amplitude $\phi_a^*$ is to be treated as a fixed {\em parameter} of the potential when we take the derivatives. The corresponding isocurvature speed is then given by
\beq
c_I^2 = {\partial\langle \pp_0 \rangle\over \partial\phi_a} \cdot \left({\partial\langle \edp_0 \rangle\over \partial\phi_a}\right)^{\!-1}\Bigg{|}_{\phi_a^*\to\phi_a}
\label{ciChain}\eeq
We have replaced straight derivatives by partial derivatives, since the time averaged quantities depend on both $\phi_a$ and $\phi_a^*$. After taking the derivative with $\phi_a^*$ fixed, we then take the limit $\phi_a^*\to\phi_a$ to obtain the correct amplitude.

With this understanding of derivatives, we arrive at a similar conclusion to the adiabatic mode: If the auxiliary pressure increases with amplitude, the isocurvature modes are stable. If the auxiliary pressure decreases with amplitude, the isocurvature modes are unstable.

\subsubsection{Application to Dim 4 Potentials}

We now illustrate this with the dimension 4 potential we examined earlier (\ref{V4}). The auxiliary potential governing the isocurvature modes can be expressed as
\beq
\Vp(\phip,\phi_a^*) = {m^2+\lambda\,\phi_a^{*2}\over 2\,\lambda\,\Tp^2}\left(1-\cos(\sqrt{2\lambda}\,\Tp\,\phip)\right)
\label{VA4s}\eeq
Earlier we plotted this potential in Fig.~(\ref{PotentialPlot}). For $\lambda>0$ this potential, as a function of $\phip$, rises more slowly than a quadratic potential; this gives rise to negative pressure. For $\lambda<0$ it rises more quickly (it can be expressed as a cosh function); this gives rise to positive pressure. So the sign of $\lambda$ determines stability, but, interestingly, in a fashion opposite to that of the adiabatic mode.

We numerically carry out the integrals of this potential to determined the time averaged auxiliary pressure $\langle\pp_0\rangle$ at some amplitude $\phi_a$ with $\phi_a^*$ held fixed. We then compute the derivatives according to eq.~(\ref{ciChain}) and then take the limit $\phi_a^*\to\phi_a$. In fact since $\phi_a^*$ only appears in the overall prefactor in eq.~(\ref{VA4s}), then its value cancels out of the ratio that gives $c_I^2$. The resulting squared speed $c_I^2$ is plotted in the lower panel of Fig.~\ref{SoundSpeedPlot}. In the figure we see that for $\lambda>0$ the squared speed $c_I^2<0$ which implies instability, while if $\lambda<0$ the squared speed $c_I^2>0$ which implies stability. We see the complementary behavior to the adiabatic mode.

Indeed by recalling $\mu_k = \pm\,i\,c_I\,k$, this result for $c_I^2$ adequately explains the presence of the thick instability band we saw earlier in Fig.~\ref{4PanelView} for $\delta\phi_\perp$ and $\lambda>0$.

\subsubsection{Application to Power Law Potentials}\label{PurePowerIso}

For pure power law potentials of the form given earlier in eq.~(\ref{Vpower}), we can determine the isocurvature speed $c_I^2$.
We do not have closed analytical forms for the auxiliary potential $\Vp$ for an arbitrary power $\ex$. However, some special cases are worth mentioning. For $\ex=2$ we find that $\Vp$ is a cosine (as discussed earlier with an additional mass term), for $\ex=1$ we just recover the quadratic potential, and for $\ex=1/2$ we find that $\Vp$ is a rational function of $\phip$. For $\ex$ non-integer, we of course need other operators to come into play around $\phi=0$; we refer the reader back to the discussion surrounding eq.~(\ref{toy}) for this issue.

Carrying out the procedure numerically, leads to the value of $c_I^2$ as a function of $q$ plotted in the lower panel of Fig.~\ref{SpeedsPowerLaw}. We see that for $\ex>1$ there is instability, while for $\ex\leq1$ there is stability. This is precisely opposite that of the adiabatic mode, whose result is plotted in the upper panel. So this proves that: {\em for an entire family of potential functions, the stability/instability of the adiabatic/isocurvature modes at long wavelengths are complementary}. The two behaviors only agree at the trivial point $\ex=1$, which is just a free theory.

\section{Circular Motion for Background} \label{Circular}

In the previous sections we studied background fields that evolved {\em radially} in field space. These radial trajectories are an attractor solution for inflation and so are strongly motivated. However, there is another class of background solutions that is worthy of study. This is when the background evolves {\em circularly} in field space. For a generic potential, this is the one other form of trajectory that will be periodic. 

Circular motion also has some physical motivation. In the previous section, we showed that under certain conditions, namely when the pressure associated with the auxiliary potential is negative, there are unstable isocurvature modes around the background radial motion. This means that the field tends to evolve locally in an angular fashion in field space. For a complex field (two field) this means either clockwise or anticlockwise motion at least locally. When such motion is established, it is important to analyze its stability. For now we treat this clockwise or anticlockwise field as homogeneous and perturb around it, even though generally it would have some spatial structure. This is relevant to the production and stability of so called Q-balls \cite{Coleman:1985ki}.

\subsection{Background Evolution}

For definiteness lets focus on two fields and organize them into a complex scalar $\phi=(\phi_1+i\,\phi_2)/\sqrt{2}$. 
Now in order to describe a potential that only depends on the magnitude $|\phi|$ it is convenient to introduce the magnitude as 
$\phim=\sqrt{2}\,|\phi|$. Using the chain rule, the equation of motion for the background is 
\beq
\ddot\phi_0 +{V'(\phim_0)\over\phim_0}\phi_0 = 0
\eeq
For circular motion, we have $\phim_0(t)=\phi_a$; a constant amplitude. Then the equation of motion becomes the equation of a simple harmonic oscillator (we assume $V'(\phi_a)>0$) with solution
\beq
\phi_0(t)={\phi_a\over\sqrt{2}}e^{-i\omega_0 t}
\eeq
(the factor of $1/\sqrt{2}$ is convenient when switching to complex notation). Here the frequency of the circular orbit is the constant
\beq
\omega_0^2={V'(\phi_a)\over\phi_a}
\label{omega0}\eeq
This is an exact closed form solution for any potential.

\subsection{Full Floquet Result}

We expand the field around the background as $\phi = \phi_0+\delta\phi$ and work to linear order as usual. The linearized equation of motion for the perturbations is
\beq
\ddot{\delta\phi}+k^2\delta\phi+{\partial^2  V\over\partial\phi_0\partial\phi_0^*}\delta\phi+{\partial^2  V\over\partial\phi_0^{*2}}\delta\phi^*=0
\label{dphiComplex}\eeq
Evaluating the coefficients on the background solution gives
\bea
\label{CircCoeff1}{\partial^2  V\over\partial\phi_0\partial\phi_0^*} &=& {1\over2}\left[V''(\phi_a)+{V'(\phi_a)\over\phi_a}\right]\\
\label{CircCoef2}{\partial^2  V\over\partial\phi_0^{*2}} &=& {e^{-2i\omega_0t}\over2}\left[V''(\phi_a)-{V'(\phi_a)\over\phi_a}\right]\,\,\,\,
\eea
We see that while the first coefficient is constant in time, the second coefficient carries periodic time dependence. 

The periodicity of the coefficient (\ref{CircCoef2}) implies that (\ref{dphiComplex}) is a type of Hill's equation. However, the time dependence in this case carries a very special structure. Since it is an exponential, we can completely remove all time dependence in the equation of motion for the perturbations by introducing the new field
\beq
\delta\psi=e^{i\omega_0t}\,\delta\phi
\eeq
The equation of motion for $\delta\psi$ is found to be
\beq
\ddot{\delta\psi}-2i\omega_o\dot{\delta\psi}+k^2\delta\psi+\!\left[V''(\phi_a)-{V'(\phi_a)\over\phi_a}\right]\!{\delta\psi+\delta\psi^*\over2} = 0
\eeq
We see that all coefficients are now time independent, so this can be readily solved. 

This can be rewritten as a collection of 4 first order differential equations. To do so, lets decompose $\delta\psi$ into real and imaginary parts $\delta\psi = (\delta\psi_1+i\,\delta\psi_2)/\sqrt{2}$ and lets introduce the momentum conjugate as $\delta\pi_1=\dot{\delta\psi}_1$ and $\delta\pi_2=\dot{\delta\psi}_2$. The system of equations can then be written as the following matrix equation for $\delta\psi_1$ and $\delta\psi_2$
\beq
{d\over dt}\left(\begin{array}{c}
\delta\psi_1\\
\delta\psi_2\\
\delta\pi_1\\
\delta\pi_2
\end{array}\right)=
\left(\begin{array}{cccc}
0 & 0 & 1 & 0\\
0 & 0 & 0 & 1\\
\bet & 0 & 0 & -2\omega_0\\
0 & -k^2 & 2\omega_0 & 0
\end{array}\right)
\left(\begin{array}{c}
\delta\psi_1\\
\delta\psi_2\\
\delta\pi_1\\
\delta\pi_2
\end{array}\right)
\eeq
where
\beq
\bet \equiv -k^2-V''(\phi_a)+{V'(\phi_a)\over\phi_a}
\eeq
The eigenvalues of this matrix are the Floquet exponents $\mu_k$. The 2 pairs of eigenvalues are found to be
\beq
\mu_k = \sqrt{-W(\phi_a)-k^2\pm\sqrt{W(\phi_a)^2+4 {V'(\phi_a)\over\phi_a} k^2}}
\label{muCircular}\eeq
where 
\beq
W(\phi_a)\equiv {1\over2}\left[V''(\phi_a)+3{V'(\phi_a)\over\phi_a}\right]
\eeq
So this provides an exact analytical result for the Floquet exponent for any potential.

\begin{figure*}[t]
  \includegraphics[width=\cw,height=\htb]{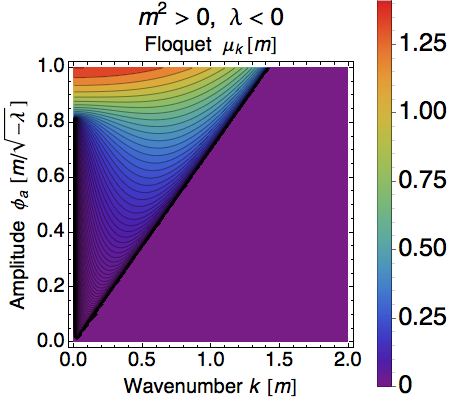}\,\,\,
  \includegraphics[width=\cw]{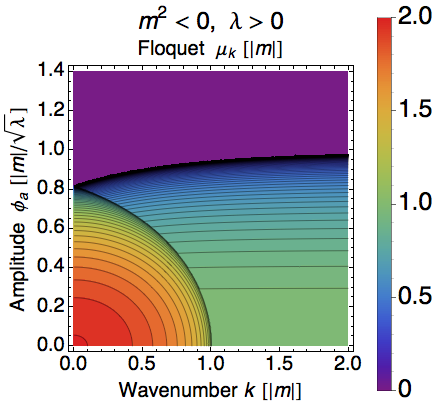}
  \caption{Contour plot of the real part of Floquet exponent $\mu_k$ for a circular background for dimension 4 potentials as a function of wavenumber $k$ and background amplitude $\phi_a$. In the left panel $m^2>0$ and $\lambda<0$. In the right panel $m^2<0$ and $\lambda>0$. 
We have plotted $\mu_k$ in units of $|m|$, $k$ in units of $|m|$, and $\phi_a$ in units of  $|m|/\sqrt{|\lambda|}$.}  
\label{CircularPlot}\end{figure*}

We evaluate this for the dimension 4 potentials as examined earlier. A plot of the result for the Floquet exponent is given in Fig.~\ref{CircularPlot}.
We have taken the upper ``+" sign in $\mu_k$, as we find this is the only exponent that can carry an instability. In the left panel $m^2>0$ and $\lambda<0$. In the right panel $m^2<0$ and $\lambda>0$. We also note that if $m^2>0$ and $\lambda>0$ there is no instability. We have plotted $\mu_k$ in units of the mass $|m|$, rather than Hubble $H$. The reason for this choice is that a circular trajectory for $\phi_0$ will only occur locally, so there is less motivation to compare $\mu_k$ to $H$.

We see that there is at most only the first instability band. This makes sense from the point of view of the quantum theory in Part 2 \cite{Part2}, where we show that higher bands are associated with annihilations $2\phi+2\bar{\phi}\to\phi+\bar{\phi}$. Since the background here is circular, it can be thought of as a collection of only particles (or only antiparticles) rather than a mixture. Hence the conservation of particle number prevents annihilations from occurring. This means an instability can only arise from particle {\em scattering}, which requires a single instability band starting at small $k$.

\subsection{Long Wavelength Limit -- Sound Speed $c_S$}

We also note a peculiar feature of the instability band in Fig.~\ref{CircularPlot}. For $m^2>0$ we see that at high amplitudes the Floquet exponent is non-zero even in the $k\to 0$ limit. For $m^2<0$ we see this behavior at low amplitudes. Let us now perform the long wavelength stability analysis to determine where this occurs and its physical explanation.

Firstly, let us discuss the energy density perturbation and the number density perturbation in this context. For a complex field, it is straightforward to show that these density perturbations are given by
\bea
\delta\nd &=&  \sqrt{2}\,\phi_a\! \left(\omega_0(\delta\psi+\delta\psi^*)+{i\over2}(\dot{\delta\psi}-\dot{\delta\psi^*})\right)\\ 
\delta\ed &=& \omega_0\,\delta\nd
\eea 
So we find that these perturbations are, at linear order, proportional to one another. Earlier in this paper, when we studied perturbations around a radial inflaton background, we found that these two kinds of perturbations were linearly independent, and that their stability charts were complementary. But when we expand around a circular background we find this new behavior. This is simple to understand as follows: We again recall that since the background is circular, it can be viewed as purely a collection of particles (or purely antiparticles) rather than a mixture. So in some sense, we only have a single type of species available, which prevents a standard type of isocurvature behavior. This means that $\delta\nd$ by itself no longer describes an isocurvature perturbation. Instead $\delta\nd\neq0$ is now associated with an adiabatic mode. There can of course still be a kind of isocurvature mode, defined by $\delta\ed=0$ (so $\delta\nd=0$ too), but these are non-resonant.

This means that there is really only one important speed that governs the behavior at long wavelengths (since the perturbations $\delta\ed$ and $\delta\nd$ have the same form). This is the sound speed $c_S^2$ of the adiabatic mode associated with pressure, as we showed earlier in Section \ref{LongWavelength}. For circular motion, the energy density and pressure of the background are given by
\bea
\ed_0 &=& {1\over2}\phi_a^2\omega_0^2+V(\phi_a)\\
\pr_0 &=& {1\over2}\phi_a^2\omega_0^2-V(\phi_a)
\eea
These are time independent, so there is no need to perform the time averaging of the earlier sections. By recalling that the squared sound speed $c_S^2$ is given by taking derivatives according to eq.~(\ref{csChain}), and by eliminating $\omega_0^2$ using eq.~(\ref{omega0}), we obtain the following analytical result for the sound speed
\beq
c_S^2 = {\phi_a V''(\phi_a)-V'(\phi_a)\over \phi_a V''(\phi_a) + 3 V'(\phi_a)}
\eeq
For example, if we apply this result to the dimension 4 potentials, we obtain
\beq
c_S^2 = {\lambda\,\phi_a^2\over2\,m^2+3\,\lambda\,\phi_a^2}
\eeq

This result can also be obtained directly from the Floquet exponent in eq.~(\ref{muCircular}) by taking the small $k$ limit then using $\mu_k = \pm\,i\,c_s\,k$. 
However this only works if $W(\phi_a)>0$. If $W(\phi_a)<0$ then $\mu_k$ does not vanish when $k\to 0$, even for the upper ``+" sign in eq.~(\ref{muCircular}). Instead the Floquet exponent approaches a $k$-independent value $\mu_k\to\sqrt{2|W(\phi_a)|}$ in this limit. So in the $W(\phi_a)<0$ regime, the field exhibits a catastrophic instability, since even perturbations that are themselves homogeneous cause large instability.

To explain this feature of instability even for homogeneous perturbations, lets analyze the condition $W(\phi_a)<0$ more closely. Suppose we were studying a pure power law $V(\phi)= {\lambda\over 2\ex}\phi^{2\ex}$. The catastrophic instability condition $W(\phi_a)<0$ implies
\beq
\ex < -1
\eeq
Now this is not normally a regime of much interest in field theory, though it may be relevant at large field values in special effective field theories. However, as we are currently probing {\em homogeneous} perturbations, the problem has essentially been reduced to a central force problem of a point particle, with distance from the origin given by $R(t) \propto |\phi(t)|$. So this condition says that central potentials with inverse power laws steeper than $V\sim -1/R^2$ are highly unstable. Indeed if one perturbs around a circular orbit, one either finds a particle trajectory that spirals out to infinity or spirals in to the origin. On the other hand, for potentials that are less steep than $V\sim -1/R^2$ (such as the classic $-1/R$ potential of Newtonian gravity) perturbations do not spiral away. The physical reason $-1/R^2$ is special is because it is competing with the energy coming from angular momentum, which itself scales as $+1/R^2$. Hence in order to have stability of the ``effective potential" one needs $\ex>-1$. If we consider more general potentials than just power laws, the generalized criteria for stability of particle orbits is $W(\phi_a)>0$.

For dimension 4 potentials, the critical value $W(\phi_{crit})=0$ occurs for
\beq
\phi_{crit}=\sqrt{-{2\,m^2\over3\,\lambda}}
\eeq
This is precisely the critical value seen in Fig.~\ref{CircularPlot}. In the left panel, with $m^2>0$ and $\lambda<0$, the catastrophe occurs for $\phi_a>\phi_{crit}$. While in the right panel, with $m^2<0$ and $\lambda>0$, the catastrophe occurs for $\phi_a<\phi_{crit}$. In the latter case, it requires that the background field is orbiting on the ``inner" part of the Mexican hat potential $\phi_a<\phi_{vev}$. Note that earlier in Fig.~\ref{NegativeMass} we only plotted $\phi_a$ on the ``outer" part of the Mexican hat potential $\phi_a>\phi_{vev}$, as the radial oscillations meant that it was redundant to include the inner part separately. While for circular orbits, these two regions are physically different.

In the case of most physical interest for us is $m^2>0$. In this case the instability in the circular background, that occurs when $\lambda<0$, suggests a type of collapse instability. This can lead to the formation of so called Q-balls \cite{Coleman:1985ki}. These are aptly named since the global $U(1)$ symmetry ensures a conserved particle number (or charge $Q$) associated with these field lumps. We note that to efficiently produce Q-balls after inflation is slightly complicated. Firstly, inflation establishes {\em radial} motion for the background. In order to obtain significant production of particle regions and separate antiparticle regions, we would like the isocurvature instability to be active, as discussed earlier. This requires $\lambda>0$. Then we would like to examine the fate of these regions. However, they will not lead to Q-balls, as this requires $\lambda<0$ for the collapse instability to occur. Instead one can imagine Q-balls forming from $\lambda<0$, even though the initial breakup of the inflaton will be towards over densities comprising both particles and antiparticles, i.e, adiabatic perturbations. This means that in simple models with $\lambda<0$, Q-balls can form, but not as efficiently as one might have naively thought otherwise. In more complicated potentials, one could imagine making the isocurvature instability active right after inflation ends, breaking up the field to particle regions and separate antiparticle regions. Then for smaller field amplitudes, having the adiabatic instability active on each of these regions, leading to the formation of Q-balls. This would presumably be highly efficient, although perhaps fine tuned. We also note that for both single or multi-field models, related structures can form, known as oscillons \cite{Amin:2010dc,Gleiser:2011xj,Amin:2011hj,Lozanov:2014zfa}; although, unlike Q-balls, they can annihilate away \cite{Hertzberg:2010yz}.

\section{Conclusions} \label{Conclusions}

In this paper we have presented Part 1 of a theory of self-resonance after inflation. For multiple fields with an internal symmetry, we have shown that the post-inflationary modes decompose into adiabatic and isocurvature modes, with long wavelength modes exhibiting a gapless spectrum as required by the Goldstone theorem.

We proved general results on the stability/instability of long wavelength modes. We constructed a sound speed from time averaging the background oscillations leading to a time averaged pressure. This time averaging is a form of coarse graining and is required to build the effective theory governed by the Goldstone modes. For positive couplings $\lambda>0$ the pressure for the adiabatic mode is positive and there is stability, while for negative couplings $\lambda<0$ the pressure for the adiabatic mode is negative and there is instability. For the isocurvature modes, we developed for the first time an ``auxiliary" potential whose time averaged pressure governs its behavior. We found that the stability structure was essentially the opposite that of the adiabatic modes. So for the classic $\lambda>0$ type of inflation modes with multiple fields, there is large resonance in the isocurvature modes, while there is very little resonance in the single (adiabatic) mode for single field models.

We mainly studied radial motion of the background inflaton field, but also considered circular motion as may arise locally in some regions after inflation. In this other limit, we were able to compute the evolution and Floquet exponents analytically in closed form. This is relevant to the possible production and stability of Q-balls. We identified a regime of catastrophe, where we saw instability even in the $k\to 0$ limit, and we explained this as related to well known results of central forces.

One of the central consequences is that there is necessarily an enhancement of power due to these various instabilities, even in regimes where it was usually unexpected; namely for $\lambda>0$. These scales do approach the horizon at early times. It would be of interest to consider any possible observational consequences of this. Ordinarily these scales are far too small for direct detection, but they may play a role in the generation of gravitational waves \cite{Zhou:2013tsa,Bethke:2013aba} or some other astrophysical phenomena. 

Altogether we presented an important step towards a complete theory of self-resonance after inflation in single and multi-field models.  For long wavelengths the behavior is determined by the physical variables pressure and auxiliary pressure. 
The Goldstone theorem organizes the adiabatic and isocurvature modes, proving that the spectrum is gapless. We believe this is the first time that the Goldstone theorem has been used in the context of self-resonance after inflation.

It is also of great interest to have a detailed understanding of self-resonance from the underlying description of the quantum mechanics of many particles.
We do this in Part 2 \cite{Part2}. This includes understanding the long wavelength phenomena using nonrelativistic quantum mechanics, the shorter wavelength phenomena using Feynman diagrams, and the explicit quantization around the classical background. Furthermore, we explore a small breaking of the symmetry, as is essential to some models of baryogenesis \cite{Hertzberg:2013jba,Hertzberg:2013mba}.

A direction for future work is to remove the internal symmetry of the Lagrangian. It would be interesting to see how this alters the structure of the various modes, and whether some analogous (``auxiliary") pressure arguments could be developed. Another possibility is to go beyond linear theory and apply these arguments to develop a theory of nonlinear fluid dynamics.

\bigskip

\begin{center}
{\bf Acknowledgments}
\end{center}
We thank Jacovie Rodriguez for help in the early stage of this project. 
We also thank Alan Guth, David Kaiser, Mustafa Amin, and Kaloian Lozanov for useful discussions.
We would like to acknowledge support by the Center for Theoretical Physics and the Undergraduate Research Opportunities Program at MIT. 
This work is supported by the U.S. Department of Energy under cooperative research agreement Contract Number DE-FG02-05ER41360. 
JK is supported by an NSERC PDF fellowship.

\section{Appendix -- Non-Canonical Kinetic Terms} \label{NonCanonical}

Here we consider a more general form of the action for scalar fields. Firstly, we focus on a single scalar field, but allow for higher derivative interactions. Secondly, we focus on just the standard two-derivative action, but allow for multiple fields with a non-trivial metric on field space.

First let us consider a single scalar field. Earlier we had truncated the action to just two-derivatives. Here we allow for higher derivatives in the scalar sector of the theory. The most general form for the action can be written as
\beq
S=\int d^4x\sqrt{-g}\left[{\mpl^2\over2}\mathcal{R}+K(\kin,\phi)+\ldots\right]
\label{Kess}\eeq
where $\kin\equiv-{1\over2}(\partial\phi)^2$. This defines a so-called K-essence model (the canonical case corresponds to $K = \kin -V(\phi)$). An example of this is DBI inflation \cite{Alishahiha:2004eh}. Interesting work on preheating in these non-canonical models includes Refs.~\cite{Karouby:2011xs,Zhang:2013asa,Child:2013ria,Bouatta:2010bp}.
In (\ref{Kess}) the dots indicate higher order gravity corrections, such as $\mathcal{R}^2,\,\mathcal{R}(\partial\phi)^2$, etc. We will ignore those corrections in this analysis. 

As we showed in Section \ref{LongWavelength}, the existence of an instability at long wavelengths is determined by a sound speed associated with time averaging the background pressure. The background pressure and density are given from the stress tensor of the scalar field. We find
\bea
p_0 &=&  K(\kin_0,\phi_0)\\
\ed_0 &=& 2{\partial K(\kin_0,\phi_0)\over \partial \kin_0}\kin_0 - K(\kin_0,\phi_0)
\eea
with $\kin_0={1\over2}\dot\phi_0^2$.
Then by time averaging over a cycle of oscillation, we obtain the sound speed $c_S^2$ as the derivative of pressure $\langle p_0\rangle$ with respect to energy density $\langle\ed_0\rangle$. This determines the Floquet exponent for small $k$, as we described in Section \ref{LongWavelength}, generalized to an arbitrary K-essence model. 

If we have multiple fields, this is still the basic methodology to construct the sound speed of the adiabatic mode. Furthermore, there may be a generalization of this result to an auxiliary pressure and energy density for the isocurvature modes, but it appears cumbersome. In the following discussion we study multiple fields, but only for the two-derivative action.

The most general two-derivative action for multiple fields involves a kinetic energy with some metric on field space $G_{ij}(\vec\phi)$. If we impose the internal rotational symmetry, this can be organized into the following form
\beq
G_{ij}(\vec\phi) = g_1(|\vec\phi|)\,\delta_{ij}+g_2(|\vec\phi|)\,\phi_i\,\phi_j
\eeq
where $g_{1,2}$ are functions of the magnitude of $\vec\phi$. In general this defines a type of so called ``nonlinear sigma model".

To be definite, lets consider the case of two fields, which we express in polar co-ordinates $\phim,\,\theta$. In this case, the most general form of the action, with the internal rotational symmetry, is
\beq
S=\int d^4x\sqrt{-g}\left[{\mpl^2\over2}\mathcal{R}-{1\over2}(\partial\phim)^2-{\kappa^2(\phim)\over2}(\partial\theta)^2-V(\phim)\right]
\eeq
where we have exploited the co-ordinate freedom on field space to express the metric in terms of a single function $\kappa(\phim)$ (the canonical case corresponds to $\kappa(\phim)=\phim$). 

For radial motion in field space, the background equation of motion for $\phim_0(t)$ is standard. Furthermore, the equation of motion for $\delta\phim$ is also the standard equation for the adiabatic modes; previously expressed as $\delta\phi_\parallel$. This can be expressed as an equation for the energy density perturbation $\delta\ed$, as given earlier in eq.~(\ref{edFric}).

For the orthogonal fluctuations (isocurvature), described here by $\delta\theta$, we find the following equation of motion
\beq
\ddot{\delta\theta}+2{\dot{\kappa}_0\over\kappa_0}\dot{\delta\theta}+k^2\delta\theta = 0
\label{thetaeom}\eeq
Now recall from Section \ref{AuxiliaryPotential}, where we studied canonical kinetic energy, that in order to pass from the adiabatic fluctuation to the isocurvature fluctuation, we needed to introduce a new field $\phip_0$, satisfying $\dot\phip_0 = \phi_0/\Tp$. This can be seen from comparing the coefficients of the first derivative terms in eqs.~(\ref{edFric},\,\ref{ndFric}). Here in eq.~(\ref{thetaeom}), we see that for the non-canonical kinetic energy, the appropriate generalization is
\beq
\dot\phip_0 = {\kappa(\phi_0)\over\Tp}
\eeq
Then by following through the methods of Section \ref{AuxiliaryPotential}, one can obtain the generalization of the auxiliary potential $\Vp$ to the case of non-canonical kinetic energy.

\end{document}